# Impact Plasma Amplification of the Ancient Mercury Magnetic Field

**Isaac S. Narrett[1]\*, Rona Oran[1], Yuxi Chen[2], Katarina Miljković[3], Gábor Tóth[2], Catherine L. Johnson[4], Benjamin P. Weiss[1]**

[1]Department of Earth, Atmospheric, and Planetary Sciences, Massachusetts Institute of Technology, Cambridge, MA, USA.
[2]Department of Climate and Space Sciences and Engineering, University of Michigan, Ann Arbor, Michigan, USA.
[3]Space Science and Technology Centre, School of Earth and Planetary Science, Curtin University, Perth, WA 6102, Australia.
[4]Department of Earth, Ocean and Atmospheric Sciences, University of British Columbia, Vancouver, BC, Canada

\*Corresponding author: Isaac S. Narrett (narrett@mit.edu)

**Key Points:**

- Giant impacts generating plasma could transiently amplify Mercury's magnetic field.
- This plasma amplified field can be recorded antipodal to the impact.
- This process can explain some elements of Mercury's remanent magnetism.

manuscript submitted to *JGR: Planets*## Abstract

Spacecraft measurements of Mercury indicate it has a core dynamo with a surface field of 200—800 nT. These data also indicate that the crust contains remanent magnetization likely produced by an ancient magnetic field. The inferred magnetization intensity is consistent with a wide range of paleofield strengths (0.2-50 µT), possibly indicating that Mercury once had a dynamo field much stronger than today. Recent modeling of ancient lunar impacts has demonstrated that plasma generated during basin-formation can transiently amplify a planetary dynamo field near the surface. Simultaneous impact-induced pressure waves can then record these fields in the form of crustal shock remanent magnetization (SRM). Here, we present impact hydrocode and magnetohydrodynamic simulations of a Caloris-size basin (~1,550 km diameter) formation event. Our results demonstrate that the ancient magnetospheric field (~0.5-0.9 µT) created by the interaction of the ancient interplanetary magnetic field (IMF) and Mercury's dynamo field can be amplified by the plasma up to ~13 µT and, via impact pressure waves, be recorded as SRM in the basin antipode. Such magnetization could produce ~5 nT crustal fields at 20-km altitude antipodal to Caloris detectable by future spacecraft like BepiColombo. Furthermore, impacts in the southern hemisphere that formed ~1,000 km diameter basins (e.g., Andal-Coleridge, Matisse-Repin, Eitkou-Milton, and Sadi-Scopus) could impart crustal magnetization in the northern hemisphere, contributing to the overall remanent field measured by MESSENGER. Overall, the impact plasma amplification process can contribute to crustal magnetization on airless bodies and should be considered when reconstructing dynamo history from crustal anomaly measurements.

## Plain Language Summary

Mercury modern global magnetic field is much weaker than Earth's, defying our understanding of planetary magnetic field generation. Orbiting spacecraft data show that Mercury's crust carries "fossil" or remanent magnetization. Understanding how this ancient magnetization was set can provide insights into Mercury's interior and evolution. As recently shown to work on the Moon, one possible explanation involves giant impacts. When a large impactor hits a planet, it can generate hot, electrically conducting vapor ("plasma") that can interact with and strengthen surface magnetic fields. Simultaneously, the impactor produces shock waves that can record this amplified magnetic field (setting this remanent magnetization). To study this, we combined modeling of large basin impacts with plasma simulations. Our results show that this process can amplify Mercury's weak field by ~10-20×. These amplified fields could have been recorded in rocks on the opposite side of planet (antipode), which could generate magnetic anomalies detectable today. Future spacecraft like BepiColombo can test this idea by measuring magnetic anomalies antipodal to large impact basins.
## 1 Introduction

The magnetometer onboard the MErcury Surface, Space ENvironment, GEochemistry, and Ranging [MESSENGER, (Anderson et al., 2007)] spacecraft revealed that Mercury possesses an internally generated magnetic field, represented by a northward-offset (from the planetary center), spin-axis aligned dipole field measuring ~200 nT at the surface equator (Anderson et al., 2011). Mercury's field presents a challenge to dynamo theory, as it is roughly two orders of magnitude weaker than that expected from core convection scaling laws (Olson & Christensen, 2006; Stevenson, 2003). Furthermore, the planet's crust contains spatially heterogeneous remanent magnetization, some of which must be the product of a past magnetic field (Johnson et al., 2018; Johnson et al., 2015; Johnson et al., 2016; Johnson et al., 2025). In particular, low altitude (<100 km) orbital passes measured strong (~10-20 nT at ~20 km) crustal magnetic fields associated with large ($10^1 – 10^2$ km) surface features at latitudes >30°N, with no low altitude passes covering the mid-latitude or southern hemisphere regions of the planet (Johnson et al., 2018; Johnson et al., 2025).

The strong crustal fields detected in the northern hemisphere are inferred to be generated from ancient (~3.9-3.7-billion-year-old, Ga) natural remanent magnetization (NRM) in Hermean surface material. Because this NRM is associated with volcanic structures, it has been proposed to be predominantly thermoremanent magnetization (TRM) acquired during slow cooling in the past field (Galluzzi et al., 2021; Hood, 2015, 2016; Hood et al., 2018; Johnson et al., 2015; Kalski et al., 2025; Oliveira et al., 2019; Plattner & Johnson, 2021). Given that both the magnetic mineralogy of Mercury's crust (Narrett et al., 2025b; Strauss et al., 2016) and evolution of the paleomagnetic field, $B_{paleo}$ (Narrett et al., 2025b) are poorly understood, these strong crustal fields can be explained by >1 km thick layers with magnetic recording efficiencies $\chi_{NRM} \sim 10^{-4} – 1$ and source paleofields $B_{paleo} \sim 0.1 – 10^1$ µT (Johnson et al., 2018). Understanding the origin of the surface magnetization is further complicated by the uncertain time-evolution of the two main magnetic field sources: (1) the modern anomalously weak dynamo-generated planetary dipole field, and (2) the stronger IMF expected for the young, faster rotating Sun (Vidotto, 2021). As such, a central puzzle in deciphering the surface magnetization is reconciling the implied strong ancient crustal fields with the weak modern planetary field. One possibility is that the past Hermean dynamo was stronger (Johnson et al., 2018; Narrett et al., 2025b).

Portions of the northern hemisphere magnetization are thought to be approximately as old as the Caloris basin formation event (Denevi et al., 2018; Fassett et al., 2012; Johnson et al., 2018; Wang et al., 2021). The Caloris basin (with a diameter of ~1,550 km and centered at ~30°N, ~190°E) is the largest confirmed impact structure on Mercury and is estimated to have formed around 3.9-3.7 Ga ago (Denevi et al., 2013; Denevi et al., 2018; Fassett et al., 2012; Fassett et al., 2009; Orgel et al., 2020; Strom et al., 2008). Impact melt, ejecta, or impact-induced effusive volcanic rocks from the Caloris event could have been magnetized as they cooled through the Curie temperature of their constituent ferromagnetic minerals over years to millions of years, thereby recording the background magnetic field as a TRM. Additionally, as shown for the Moon (Narrett et al., 2025a), large basin-forming impactors can generate ionized vapor or plasma ("impact plasma") that can amplify global dipole magnetic fields. This amplification is largest at the basin antipode and when the impact occurs near the magnetic axis of a dipole field (i.e., magnetic pole). Such fields might be recorded as TRM in igneous rocks or as shock remanent magnetization (SRM) in rocks exposed to pressure waves from the impact event (Hood, 1987; Hood & Artemieva, 2008; Hood & Huang, 1991; Hood & Vickery, 1984; Narrett et al., 2025a; Oran et al., 2020). Using two dimensional (2D) impact-physics simulations and three dimensional-magnetohydrodynamic (3D-MHD) simulations, (Narrett et al., 2025a) showed that for the ancient Moon, an impact forming an Imbrium-sized basin could amplify a ~2 µT dipole surface field to ~43 µT at its antipode around 60 minutes after the impact, provided that the impact occurs close to the magnetic pole (within ~30° latitude). Because impact-induced pressure waves would focus at the antipode at approximately the same time as the field was amplified to its maximum value, rocks at the antipode could record the amplified field as SRM. (Narrett et al., 2025a) also showed that impact ejecta landing and cooling at the antipode (>4 hours after impact) would not be able to record the short-lived (~1 hour) plasma amplified dynamo field as a TRM (Citron et al., 2025; Hood & Artemieva, 2008; Wakita et al., 2021), such that TRM found on the Moon likely recorded the dynamo field while SRM could have recorded a dipole field distorted and greatly amplified by the impact.

Mercury, along with the Moon, are the only known airless bodies whose crust contains remanent magnetization indicative of past magnetic fields. As such, Mercury is a natural laboratory for further testing the impact plasma dynamo amplification hypothesis: if such a mechanism is an important process for magnetizing planetary bodies, we should investigate its role on airless bodies beyond the Moon. Motivated by this, we assess whether plasma generated from the Caloris basin formation event could have amplified the local Hermean magnetic field and explain at least some of the Mercury surface magnetization. Specifically, we aim to answer the following questions. (1) Could impact plasma amplification of the Hermean dynamo field have magnetized the Caloris antipode? (2) Can this impact plasma process explain the strong northern surface magnetization or does this require a stronger source field? To answer these questions, we performed impact hydrocode simulations of the Caloris basin formation combined with MHD simulations of the impact plasma evolution in the near-Mercury space environs. We then coupled these simulations with crustal magnetization models to quantify the expected crustal magnetization and associated fields. We specifically focus on the circum-Caloris region and the Caloris antipode, which lies in the so-called H-11 Discovery Quadrangle [22°S-65°S and 270°E-360°E, (Trask & Dzurisin, 1984)].

This paper is organized as follows. Section 2 describes the impact simulation setup, ancient Mercury solar wind and magnetic field conditions, and the MHD model used to perform the impact plasma simulations. Section 3 presents the results of the MHD simulations. Section 4 discusses the implications of the results and calculates the possible magnetization set by this impact process. Section 5 contains the conclusions with recommendations for testing the impact plasma amplification hypothesis with future spacecraft measurements.

## 2 Methods
### 2.1 Caloris Basin Impact Simulations

We chose to model the Caloris basin event because its formation age (3.9-3.7 Ga ago) is linked to the age of some of the widespread, northern hemisphere surface magnetization in its surroundings and because it is similar in size to the Imbrium basin on the Moon where this impact process was previously studied (Hood & Artemieva, 2008; Narrett et al., 2025a; Oran et al., 2020). Such large basins will produce the largest volumes of plasma and the most energetic shock waves, both key ingredients for amplifying a dynamo field and recording it as SRM. Additionally, although we did not explicitly conduct hydrocode and MHD models of other basin forming events, we used our results from this work and previous studies to qualitatively assess the possibility of field amplification and crustal remanence produced by basins of similar size to Caloris and Imbrium.

Our impact simulations were performed with the shock physics multi-material, multi-rheology, Lagrangian-Eulerian code iSALE-2D (Amsden et al., 1980; Collins et al., 2010; Wünnemann et al., 2006) to model the formation of a Caloris-sized basin. The numerical grid cell sizes were 2.5 by 2.5 km. The impactor had a diameter of 120-km and an impact velocity of 30 km s$^{-1}$, representing the median impactor parameters for creating a Caloris-sized basin at the orbit of Mercury (Ivanov, 2008; Le Feuvre & Wieczorek, 2011). Mercury's crust and mantle were modeled with basalt and dunite equations of state, respectively, while the impactor was modeled with dunite (Potter & Head, 2017) [note that the choice of the equation of state for the crust does not make a significant difference on the results when simulating such large impact events (Benz et al., 1989; Miljkovicc et al., 2013)]. The crust and mantle on Mercury were modeled to be 40-km (Beuthe et al., 2020; Padovan et al., 2015) and 400-km (Davies et al., 2024) thick, respectively.

Two sets of such simulations were conducted: one focused on initial vapor production and flux and the other on stress propagation through Mercury's crust to assess antipodal focusing and interior stress responses. The latter also models the Caloris basin formation to completion. In both sets of simulations, we used 24 cells-per-projectile-radius (CPPR). This resolution is a nominal numerical resolution when simulating crater formation and similar to previous works on the Moon (Narrett et al., 2025a; Oran et al., 2020). Given the necessity to facilitate a numerical mesh to accommodate for vapor expansion as well as stresses throughout the entire planet, this modest CPPR value is deemed satisfactory.

The first set of simulations tracked the impact-generated vapor's density, temperature, and speed which are subsequently used as boundary condition inputs to the MHD simulations. Using kinetic energy and vaporization scaling relationships for hypervelocity impacts (Ahrens & Okeefe, 1977; Hood & Huang, 1991; Miyayama & Kobayashi, 2024), the total vaporized material should include all material within ~2 impactor radii, which gives ~$2.4 \times 10^{19}$ kg. We found that the total, simulated vapor mass generated is ~$2 \times 10^{19}$ kg, in agreement with the analytical approximation.

The impact simulation was run for 1,200 s after the impact. Around 300 s after the impact, the generated vapor has a peak flux of ~$10^{16}$ kg s$^{-1}$ at 500-km above the surface; this vapor has sufficient vertical velocity to escape the forming basin walls, circumvent the ejecta curtain, and spread around the planet (Figure 1). Although there is a transient spike in the vapor flux around ~600 s due to a turbulent structure in the initial plume expansion, the overall vapor flux starts to decay starting around 1,000 s after impact, setting the total vapor (impact plasma) emission time frame as ~700 s. By the end of the 1,200 second simulation, the vapor mass has decayed by >3 orders of magnitude from its initial value. The mass density of the expanding vapor is ~8 kg m$^{-3}$, corresponding to the maximum generated vapor flux at 300 s. We use the latter mass density as the initial condition for the expanding impact plasma in the MHD simulations. After the initial ~$10^4$ K temperature generated upon impact within the initial transient crater, which allows for the vapor to be significantly ionized and treated as a plasma [(Bruck Syal & Schultz, 2015; Hood & Vickery, 1984; Oran et al., 2020), see also materials and methods of (Narrett et al., 2025a)], the freely expanding impact plasma has a net temperature of ~4,000 K. This is the temperature we use in the MHD simulations.

The second set of simulations globally tracked the effects of the Caloris basin forming event up to ~6,000 s after impact. These modeled the evolution of the impact-induced pressure waves, which converge and amplify at the basin antipode (Schultz & Gault, 1975; Watts et al., 1991). By coupling the spatial and temporal evolution of the pressure waves (which can provide the energy to shock magnetize the surface material) and magnetic field, we were able to calculate the potential SRM generated crustal fields.

## 2.2 Ancient Mercury Solar Wind and Parameter Space

To adequately model the magnetic field environment ~3.9-3.7 Ga ago, we first derive the solar wind and IMF conditions at the orbit of Mercury. The flowing solar wind can be mathematically described by the MHD approximation over scale lengths of order the Hermean radius, $R_M$, using the average fluid quantities: mass density, $\rho$, bulk velocity vector, $\boldsymbol{u}$, average temperature, $T$, and IMF vector, $\boldsymbol{B_{IMF}}$. These fluid state variables depend on the mass loss rate (MLR, the rate at which plasma is ejected from the solar surface), solar rotation rate, and surface magnetic field. Over time, stars generally lose both angular momentum and mass, meaning that the young Sun would have been more active during the time when the Hermean crustal magnetization was acquired. Analytical models, numerical simulations, and observations of young solar analogues give insight into the heightened MLR, spin rate, and surface magnetic field of the ancient Sun (Güdel, 2007; Vidotto, 2021). The observations and physics-based scaling laws for the time-evolution of MLR yields values between ~$10^{-13}$ and ~$5 \times 10^{-12}$ $M_\odot$ yr$^{-1}$ (Johnstone et al., 2015; Vidotto, 2021). The spin rate, which effects the surface magnetic field, is thought to have ranged between $2\Omega_\odot$ and $5\Omega_\odot$, where $\Omega_\odot$ is the present day rotation rate (Gallet & Bouvier, 2015; Vidotto, 2021). Based on these heightened spin rates and observations of young, Sun-like stars, we estimate that the average surface magnetic field ranged anywhere from 1 to 3 mT (Vidotto, 2021; Vidotto et al., 2014), in line with a recent catalog of young solar analogue surface magnetic fields [see Figure 8 in (Evensberget et al., 2022) for a comprehensive survey of surface magnetic fields from theory and observations].

We now translate these solar properties to the plasma conditions at Mercury. To self-consistently and accurately approximate the solar wind properties, we utilize the Weber and Davis (Weber & Davis, 1967) model for the propagation of a magnetized solar wind as implemented in (Johnstone, 2017). This model builds on the pioneering, hydrodynamic Parker solar wind model (Parker, 1958) by solving the MHD equations including the solar magnetic field and rotation rate. Using this model and the previously derived ranges of the MLR, rotation rate, and surface magnetic field, we can approximate the mass density, velocity, temperature, and IMF of the solar wind at any point in the solar system (Burlaga, 1995; Echer et al., 2020; Oran et al., 2018). Today, Mercury's semi-major axis is 0.387 AU and eccentricity, $e$, is ~0.2 (Einstein, 1915), although there is uncertainty ($0.0 \leq e \leq 0.4$) in the ancient eccentricity due to the chaotic nature of inner solar system dynamics (Laskar, 1994). This uncertainty in the eccentricity results in a range of orbital distances over the Hermean year: $0.232$ AU $\leq r \leq 0.542$ AU. In the absence of additional constraints, we assume the modern semi-major axis distance (0.387 AU) for approximating the average, ancient solar wind parameters and later discuss these parameters effect on the impact plasma amplification. For the MHD simulations, we ignore the spin-axis rotation of Mercury, as even the relatively low-likelihood 5:1 spin-orbit resonance resulting in the fastest possible spin-axis rotation (>18 days) is much longer than the impact plasma process (~hours) [(Noyelles et al., 2014; Mark A. Wieczorek et al., 2012)].

Based on the range of the MLR, solar rotation, and solar magnetic field, the possible ranges of the ancient solar wind parameters are: $10^2$ amu cm$^{-3}$ $\leq \rho \leq 10^3$ amu cm$^{-3}$, $u$ ~ 600 - 1200 km s$^{-1}$, $T$ ~ $10^5$ - $10^6$ K, and 300 nT $\leq B_{IMF} \leq$ 600 nT (Tables 1 and 2). The overall magnetic field amplification calculated by our simulations is not sensitive to the exact choice of upwind solar wind parameters within these ranges because the impact plasma thermal pressure is >13 orders of magnitude larger than the solar wind dynamic pressure (Narrett et al., 2025a). As previously shown in the case of the impact plasma amplification of the IMF and ancient weak lunar dipole field (Narrett et al., 2025a), the main driver of any magnetic field amplification is the initial, unperturbed field strength and direction relative to the impact location. Furthermore, the magnitude of the magnetic field amplification is approximately proportional to the initial average field, as shown in the previous findings [see supplementary material in (Narrett et al., 2025a)]. This means that modifying the IMF and dipole field strength will result in a roughly proportional scaling for the maximum amplified field. Thus, we use the following baseline solar wind parameters: $\rho = 1500$ amu cm$^{-3}$, $u = 600$ km s$^{-1}$, $T = 3 \times 10^5$ K, and $B_{IMF} = 500$ nT. We perform additional simulations to show how changing the IMF and dipole field magnitude and direction affects the overall magnetic field amplification and subsequent crustal field generation (varying the impact location and IMF geometry also results in different initial antipodal field strengths due to the magnetospheric field dependence on solar wind and IMF conditions). We also note that these estimations for the solar wind conditions at ancient Mercury are similar to previous estimates (Heyner et al., 2012).

Given the poorly constrained properties of the ancient surface magnetization (Section 1), there is greater uncertainty with the strength of the ancient dynamo. As such, our baseline simulations start with the modern dipole field strength and geometry, a spin-axis aligned field with a strength of ~200 nT at the surface of the magnetic equator (Anderson et al., 2011). Using the parameters in Tables 1 and 2, we define six cases to simulate. Cases 1 and 2 were chosen to show the basic impact plasma process with the impact antipode surface normal, $\hat{n}_i$, perpendicular and parallel to the IMF. Cases 3, 4, 4-Reverse, and 5 were chosen to show the effects of shifting the spin-axis aligned dipole center by 20% northward along the spin axis [following estimates of the approximated dipole model from MESSENGER data (Anderson et al., 2011)], translating the impact to occur at the Caloris latitude 30°N, and changing the IMF magnitude and direction. In particular, Case 5 represented the approximated likely average IMF configuration for ancient solar conditions from the Weber and Davis (Weber & Davis, 1967) model (Johnstone, 2017), (we calculated the average $B_{IMF, X}$ and $B_{IMF, Y}$ components, which are the radial and azimuthal components in the Sun-centered ecliptic plane, respectively). These components are likely dominant relative to the $B_{IMF, Z}$ (out-of-ecliptic plane component), as seen in modern IMF measurements throughout the solar system (Burlaga et al., 2002; Chang et al., 2022; Echer et al., 2020; James et al., 2017). Based on the range of possible ancient solar rotation rates ($2\Omega_\odot$ and $5\Omega_\odot$), we found that the Parker spiral angle [angle between $B_{IMF, X}$ and $B_{IMF, Y}$ components (Parker, 1958)] at Mercury likely ranged between ~30° and ~50°, meaning that $B_{IMF, X}$ and $B_{IMF, Y}$ are similar in magnitude on average. Lastly, Case 6 assessed the possibility of generating stronger crustal fields by increasing the Hermean dipole field strength by 10✕ (2 µT), in line with estimates of a stronger, ancient dynamo (Narrett et al., 2025b).

## 2.3 MHD Simulation Setup
We performed the 3D-MHD simulations in two steps: (1) we calculated the interaction between the ancient solar wind and the Hermean dipole field until it reached steady-state, and then (2) incorporated the impact simulation-derived impact plasma (see Section 2.1) into the domain via a time-dependent boundary condition, letting the system evolve until the magnetic field amplification relaxed towards its original state.

The 3D MHD simulations were performed with the Block-Adaptive Tree Solarwind Roe-type Upwind Scheme (BATS-R-US) code (Powell et al., 1999; Tóth et al., 2012). The simulations solved the ideal and resistive MHD equations [equations 1-4 in (Narrett et al., 2025a)]. The ideal MHD equations were solved outside the Mercury surface, whereas inside the body, the plasma velocity and density were set to zero, solving the resistive induction equation for the magnetic field evolution with a radial internal conductivity profile as described in (Jia et al., 2015; Jia et al., 2019; Li et al., 2023), accounting for Mercury's large core. The equations were solved on a spherical grid with outward logarithmically-scaled cell size in the radial direction, allowing for higher resolution in the planet surface and interior relative to the planet's surroundings. The computational domain extended from radius $0.8R_M$ to $20R_M$ (outer boundary) to fully allow for the magnetospheric cavity to form (Supporting Information Figure S1). The grid resolution was increased (>9× smaller volume cells) with BATS-R-US's mesh-refinement capabilities for the cells within $3R_M$. The inner boundary was set at the top of the highly conducting Hermean core (radius of $0.8R_M$). The radial cell size at the Mercury-centered radial distance, $R = 0.8$, $1.0$, and $20\ R_M$ is $0.008$, $0.011$, $0.9\ R_M$, respectively, while the angular cell size at $R = 0.8$, $1.0$, and $20\ R_M$ is $0.012$, $0.016$, and $1.2\ R_M$. The cell resolution at the surface and in the interior was similar to that of previous Mercury simulations performed with BATS-R-US, capturing the relevant magnetospheric and core induction physics (Jia et al., 2015; Jia et al., 2019; Li et al., 2023). Additionally, this grid was shown to fully capture the resolved antipodal magnetic field amplification for similar solar wind/impact plasma regimes and magnetic field strengths (see supplementary material in (Narrett et al., 2025a) for grid convergence discussion) as seen in this study. Lastly, the magnetic field amplification resulting from each simulation case (see below) agreed well (<10% difference) with previous findings (Narrett et al., 2025a) for approximating the maximum amplified field (see Sections 2.4 and 3), giving confidence that this grid has sufficient cell resolution.

The polar and azimuthal (angular) boundary conditions of all quantities are periodic along the polar axis and zero meridian. At the core-mantle boundary, the boundary conditions were set to the constant dipole value and any perturbations to the magnetic field are reflected as the core can be approximated as a perfectly conducting sphere (Narrett et al. 2025a). The Hermean surface is not a boundary to the magnetic field; rather, the induction-diffusion equation transitions from the regime of the plasma's perfect conductivity outside the body to the finite conductivity regime for the crust and mantle. At the surface, the inflowing solar wind was absorbed and any outflow was inwardly reflected, as the Hermean surface is not a significant source of plasma. The impact plasma was emitted from the impact basin and was only permitted to have tangential velocity with respect to the surface. At the outer edge of the domain, the boundary conditions for the solar wind were set to inflow or outflow based on the direction of the solar wind velocity relative to the boundary normal vector.

To mitigate the violation of the divergence free condition of the magnetic field from numerical discretization errors in the solution of the induction equation, we employed both the eight-wave and the hyperbolic cleaning methods (Powell et al., 1999; Tóth et al., 2012). We used a semi-implicit time-integration to solve the magnetic induction equation inside the body, while outside the body, we used an explicit scheme with Courant number 0.8 (Tóth et al., 2012). Several adjustments were required to solve the time-accurate (Tóth et al., 2012; van der Ven et al., 1997) evolution post-impact due to the constraints of computational power along with the accuracy and simulation time needed. The post-impact time-accurate coupled evolution of the plasma and magnetic field were solved to second-order accuracy with a Courant number of 0.4. To increase the allowable explicit timestep, we applied the so-called "Boris correction" factor (Boris, 1970; Tóth et al., 2012) of 0.02 to artificially reduce the speed of light, which is a limiting factor of the MHD fast wave speed. Finally, to allow for the solution to propagate across the computational poles, we reduced the order of accuracy of the numerical scheme to first order in space in the cells immediately surrounding the computational pole. These numerical changes were previously (Narrett et al., 2025a) verified to not significantly affect the resultant magnetic field amplification process.

## 2.4 Predictions for Impact Plasma Amplification for Mercury

We next review the results of a previous study of impact amplification of the lunar dynamo (Hood & Artemieva, 2008; Narrett et al., 2025a; Oran et al., 2020). Although this process was studied for lunar impacts, the present study is distinct in several ways: we consider a Mercury-sized body, differing solar wind conditions, a unique planetary dipole field strength and geometry, and impacts with the size and location of the Caloris basin event. Previous studies [27, 28] showed that the highly conducting impact plasma transports, compresses, and amplifies the local magnetic field towards the antipode. As the plasma expands around the surface, it compresses the initial, unperturbed magnetic field into a smaller volume in the antipodal region, such that the reduction in surface area through which the field threads leads to field amplification [from the frozen-in flux theorem (Gombosi, 1998)]. Simultaneously, magnetic energy is lost through ohmic dissipation in the crust, such that the flux transport across the body is intrinsically coupled with dissipation. This dissipation is nonlinear with respect to magnetic field and requires numerical modeling, as simple analytical flux-conservation arguments would overestimate the surface amplification factor. The expansion of the impact plasma is determined almost completely by the thermal plasma pressure and planetary gravity field, while negligibly affected by the local dipole field, IMF, and solar wind. Although the maximum magnetic field occurs in the antipode, where the impact plasma converges, other magnetic field enhancements are seen in locations where the impact plasma compresses the local magnetic field within the planetary surface. This compression occurs in areas where the local magnetic field is tangential to the surface, as surface currents form at this boundary to exclude the internal magnetic field from entering the highly conducting impact plasma. In turn, these currents amplify the magnetic field by ~3✕ at and just below the surface, analogous to the Chapman-Ferraro currents amplifying the planetary dipole field at the magnetopause (Chapman & Ferraro, 1931; Narrett et al., 2025a; Schield, 1969).

One of the key findings from (Narrett et al., 2025a) was that the direction of the planetary (dipole) magnetic field at the surface near the impact location and antipode (in other words, the magnetic latitude of the impact site) governed the overall amplification of the antipodal magnetic field (Figures 2 and 3). These simulations (Narrett et al., 2025a) of the impact plasma expansion in the presence of a $10^3$ nT equatorial surface dipole field (with similar results for $10^2$ nT and $10^4$ nT) showed that impacts occurring near the magnetic pole, where $\boldsymbol{B} \parallel \hat{\boldsymbol{n}}_i$ (and $\hat{\boldsymbol{n}}_i$ is the normal to the surface at the impact antipode) resulted in maximized magnetic field amplification at the antipode surface (~21✕ initial local field), as parallel field lines are compressed together. Conversely, impacts occurring near the magnetic equator ($B \perp \hat{n}_i$) resulted in the least overall magnetic field amplification at the antipode surface (~6✕ initial local field) due to the compression of antiparallel field lines, which created a region of low-field occupying the compressed volume. Impacts occurring at locations between the dipole field pole and equator produced intermediate amplification factors, with the amount governed by the relative proportion of parallel and anti-parallel field compressed in the antipode. The overall magnetic field amplification reached its maximum value above the antipodal surface in each case because of the low magnetic field dissipation in the highly conducting space plasma environment. By comparison, at the antipodal surface and in the interior, the resistivity of the crust and mantle (Johnson et al., 2016) inhibits both the magnitude and time duration of the magnetic field amplification.

Taking the ancient IMF together with the previous findings of amplification factors, we expect maximum antipodal surface fields from ~3 to ~19 µT for an initial, average surface field ranging from 0.5 µT from the IMF without any pileup at Mercury to 0.9 µT from the superposition of the northward shifted modern Hermean dipole and IMF. Generally, impacts for Caloris's location at ~30°N should result in maximum antipodal amplification factors of ~11✕, based on the previous tests done relating impact location relative to dipole field axis [see supplementary material of (Narrett et al., 2025a)]. For an amplification upper limit, an IMF that is dominantly perpendicular to the solar wind flow direction results in magnetic field pileup at the core-mantle boundary and can create a heightened initial average surface field (~$3B_{IMF}$, yielding maximum antipodal amplified field of 31 µT). This magnetic field pileup amplification is based on the upstream solar wind and IMF conditions and the relative interplay between the solar wind convection timescale and planetary interior structure magnetic diffusion timescale (Anand et al., 2022; Oran et al., 2018; Poppe & Fatemi, 2023).

Furthermore, as the undisturbed IMF magnitude (0.5 µT) is similar to the planetary dipole field strength (~0.2 µT), we expect that the angle between the IMF and dipole field will affect the overall amplification, because initially the solar wind is strong enough to compress the dipole field and carry the IMF to the surface. With our understanding of previous studies and our new results, the total amplification magnitude can be estimated (for comparison with each simulation in Section 3) by the initial antipodal field strength, the relative geometry between the IMF and dipole field, and the impact location with respect to the dipole magnetic field axis.

### 3 Results
### 3.1 Cases 1 and 2: Impacts with Centered Dipole from the Pole and Equator

Case 1 (Figure 4) and Case 2 (Figure 5) represent the evolution of the impact plasma from an impact at the surface equator with the body-centered Hermean dipole field and the IMF direction perpendicular and parallel to the impact antipode normal ($B_{IMF} \perp \hat{n}_i$ and $B_{IMF} \parallel \hat{n}_i$), respectively. These simulations (with the IMF dominant to the planetary field) are analogous to the scenarios where the impact plasma was launched from the lunar dipole pole and equator, resulting in the respective maximum ~21✕ and minimum ~6✕ amplification of the antipodal surface magnetic field (Narrett et al., 2025a). For both cases, the impact plasma evolves around the Hermean surface, convecting the local magnetic and reaching the antipodal region within ~30 minutes. At around ~38 minutes after impact, the magnetic field is maximized in both Cases 1 (~2.8 µT) and 2 (~11 µT), matching the predicted ~6✕ and ~21✕ amplification of the initial ~500 nT IMF seen for the analogous equatorial and polar lunar dipole impacts [(Narrett et al., 2025a), see Section 3.2]. These heightened antipodal fields are short-lived, decaying back to the initial strengths after ~20 minutes due to ohmic dissipation (Narrett et al., 2025a; Oran et al., 2020) in the Mercury crust (Johnson et al., 2016).

**3.2 Cases 3, 4, 4 - Reverse, 5, and 6: Impacts with Shifted Dipole from 30°N**
To assess the unique nature of the Mercury environment and how it deviates from previous findings, the rest of the simulations were performed with the spin-axis aligned dipole center shifted northward by $0.2R_M$ (Anderson et al., 2011) and the impact plasma released from 30°N (Caloris basin center). Case 3 (Figure 6) and Case 4 (Figure 7) were designed to find the limiting scenarios of $\boldsymbol{B_{IMF}} \perp \hat{\boldsymbol{n}}_i$ and $\boldsymbol{B_{IMF}} \parallel \hat{\boldsymbol{n}}_i$ analogous to those for the centered dipole impacts in Section 3.1. As expected from the dominance of the impact plasma pressure over other forces, these simulations result in the maximum antipodal magnetic fields at around ~38 minutes after impact. With an average initial surface field of ~834 nT, Case 3 resulted in a maximum antipodal field of ~13.7 µT (~16✕ amplification factor), deviating from the ~6✕ amplification found for $\boldsymbol{B_{IMF}} \perp \hat{\boldsymbol{n}}_i$ with the body-centered dipole (Case 1). This difference can be explained by both the shifted dipole field and the relative geometry of the dipole field and IMF. The shifted dipole results in an asymmetrically distributed magnetic field between the northern and southern hemispheres. With this dipole geometry and impact location, the impact plasma is no longer compressing together symmetric volumes of anti-parallel magnetic, such that the amplified magnetic field experiences parallel field compression in the antipode. Yet, the absolute maximum amplification factor is not achieved due to the anti-parallel geometry of the dipole field ($\boldsymbol{B_{Dipole}}$) and IMF in the antipodal region (Figure 3), as $B_{IMF, X}$ and $B_{IMF, Z}$ are anti-parallel to $B_{Dipole, X}$ and $B_{Dipole, Z}$ (in the Mercury-body centered coordinate system, Table 1), respectively. Thus, the total amplification can be approximated by multiplying the average initial surface field of ~834 nT by 21 and subtracting the amplified anti-parallel antipodal dipole field (0.333 µT) for impacts at this magnetic latitude of 11✕0.333 µT (21 ✕ 0.834 µT – 11 ✕ 0.333 µT), resulting in ~13.9 µT. This agrees well with the simulation maximum antipodal magnetic field of ~13.7 µT.

Case 4 ($\boldsymbol{B_{IMF}} \parallel \hat{\boldsymbol{n}}_i$) had an initial average surface field of ~694 nT, slightly less than the initial average antipodal field in Case 3 due to $\boldsymbol{B_{IMF}}$ having a smaller component perpendicular to the flow direction (Table 1), resulting in less magnetic field pileup. Again, for this case, $B_{IMF, X}$ and $B_{IMF, Z}$ are anti-parallel to $B_{Dipole, X}$ and $B_{Dipole, Z}$, respectively, in the antipodal region, hindering the total magnitude of the amplification. As such, we calculate that the maximum surface field will be 10.8 µT (21 ✕ 0.694 µT – 11 ✕ 0.333 µT), which is in good agreement with the simulation result of 10 µT. To further demonstrate the effect of having the IMF vector anti-parallel to the surface dipole direction, we performed Case 4 – Reverse, for which we flip the upstream IMF direction by 180° in the *x-z* plane (i.e. $-\boldsymbol{B_{IMF}} \parallel \hat{\boldsymbol{n}}_i$). As expected, for Case 4 – Reverse, we found a higher maximum surface field in the antipode of 12.7 µT, due to $\boldsymbol{B_{IMF}}$ being parallel to $\boldsymbol{B_{Dipole}}$, eliminating the canceling effects from the relative geometry (Figure 3).

After showing the general amplification dependence on the impact location, IMF, and dipole field relative geometry, we ran an additional simulation to demonstrate the possible amplification for the average IMF orientation (i.e., Parker spiral orientation) based upon theory and observations. For this test (Case 5, Figure 8), we find that the maximum antipodal field reaches 13 µT. This finding agrees well with the approximation that the initial average surface field of ~668 nT should be amplified by ~21✕ to 14 µT without any negating factors from an anti-parallel dipole field [$B_{IMF, X}$ was parallel to $B_{Dipole, X}$ and there was no *Y*-component of the dipole field due to the dipole axis aligned with the spin-axis (Anderson et al., 2011)].

To further study the range of possibilities for the ancient Mercury magnetic field environment, we performed an additional simulation to show how the impact plasma amplification process would differ with a stronger dipole field (Johnson et al., 2018; Narrett et al., 2025b). From the previous study with the ancient lunar dynamo (Narrett et al., 2025a), the amplification was found to be roughly proportional (≤10% variation when changing the dipole field strength by an order of magnitude) with the initial dipole field strength for fields within 0.1 µT and 10 µT. In particular, Case 6 employed a 10✕ stronger dipole field, which resulted in an initial average surface field of ~3.3 µT. From this average surface field and impact at 30°N, we found the maximum field was ~33 µT (Figure 9), agreeing well with the predicted maximum antipodal surface field of ~34 µT (~11✕ amplification factor).

## 4 Magnetization Process for Recording Amplified Antipodal Field

Our simulations of the impact-generated plasma amplification of a weak (~200 nT equatorial field) Hermean dynamo and ambient IMF produce surface fields at the basin antipode of ~$10^1$ µT, with the amplified field lasting ~20 minutes (Figure 10). To assess whether this field can be recorded by the Caloris antipode surface material, we must assess the possible NRM acquisition mechanisms and their corresponding efficiencies and timescales.

### 4.1 Caloris-Induced Impact Pressures and Antipodal NRM

As proposed for the Moon (Citron et al., 2025; Hood & Artemieva, 2008; Wakita et al., 2021), heated impact ejecta can traverse the space environs, land and cool in the basin antipode, and possibly record the impact plasma amplified field as a TRM. However, this process was deemed to not be viable for the Moon, as the amplified field would likely have long subsided by the time of first ejecta arrival (>4 hours), even for the most preferential impact trajectory (Narrett et al., 2025a; Wakita et al., 2021). We can place a limit on the shortest ejecta arrival time at Mercury's antipode by calculating the circular orbit travel time (~43 minutes), noting that a similar estimation for the Moon underpredicts the shortest travel times found for lunar basin impact ejecta by at least a factor of four (Wakita et al., 2021). As the impact plasma amplified field has decayed by >90% at ~60 minutes, we conclude that it is extremely unlikely that the ejecta deposited in the antipode could record such a field. If Caloris ejecta were to record a TRM upon landing in the antipodal region, the magnetizing field would be a combination of the relaxed, nominal Hermean dynamo field and any local crustal remanent field. Ejecta blocks would need to be likely >10 m in scale size in order to cool and record a steady TRM [extreme ancient solar wind would erode the dayside magnetosphere and surface would be subject to non-steady, directionally varying IMF (Glassmeier et al., 2007; Hood & Schubert, 1979; Jia et al., 2019; Narrett et al., 2025b)], as the conductive cooling timescale of rocks of this size (~year) is of order the Mercury rotation period (~0.5 year). Alternatively, small rocks would need to be <1 cm in scale size to cool in ~minutes and record the time-variable (~10s of minutes) superposition of the IMF and dipole field (James et al., 2017). Future measurements of the Caloris antipode region can search for deposited ejecta and co-located magnetic anomaly sources (Rothery et al., 2020), [as has been done on the Moon (Citron et al., 2025; Wakita et al., 2021)].

Also, as proposed for the Moon (Hood & Artemieva, 2008), the local antipodal surface material can be shocked by the converging pressure waves from the basin forming impact event. Previous studies for Mercury suggest that pressure waves from a Caloris-sized event can reach 0.1-2 GPa near the antipode up to ~60 minutes after impact (Hughes et al., 1977; Lü et al., 2011; Schultz & Gault, 1975; Watts et al., 1991). These pressure waves can produce SRM, with the amplified impact plasma field being nearly instantaneously recorded over the pressure deposition timescale. To estimate whether sufficient pressures existed concurrently with the amplified field, we first calculated the maximum pressure experienced at each location inside the antipodal region during times when the amplified field was >10 µT (Figure 11). Then, we averaged this spatial map of maxima over the entire region, finding this to be ~0.4 GPa. These pressures were not exceeded at later times when the amplified field decayed, implying that this SRM could survive the impact event.

To determine whether these impact-induced pressure waves could magnetize the surface material and generate measurable crustal fields, we modeled these magnetic field structures at plausible spacecraft altitudes (Figs. 12 and 13). As there are currently no sensitive crustal field measurements taken in the southern hemisphere, we explored the magnetized material volume, SRM efficiency ($\chi_{SRM}$), and paleofield ($B_{paleo}$) parameter space required to produce ~5 nT signals at 20-km altitude by future spacecraft [e.g., BepiColombo Mercury Planetary Orbiter [MPO], (Heyner et al., 2021; Rothery et al., 2020)]. A 5 nT field is taken as a reasonable estimate of a signal readily detectable in spacecraft observations, noting that the crustal field strength is proportional to the magnetized volume and $\chi_{SRM}$ and can be accordingly scaled to estimate the required parameter space to create the strong ~10-20 nT fields or any other discernable signal of given strength. The magnetized volume was represented by uniformly magnetized cylindrical disks of radius 20-km (chosen to maximize the field given the 20-km altitude spatial scale), with varying thicknesses (Caciagli et al., 2018; Kalski et al., 2025; M. A. Wieczorek et al., 2012). We estimate an upper limit of $\chi_{SRM}$ values for different surface materials using the TRM efficiencies, $\chi_{SRM} = 0.1\chi_{TRM}$, in (Kletetschka et al., 2006; M. A. Wieczorek et al., 2012) and with $\chi_{SRM} \leq 0.1\chi_{TRM}$ (Figure 13). The magnetic mineralogy of the Hermean surface is poorly constrained, with measurements of the average crustal (~1.75 wt%) Fe content and compositional features indicating formation in extremely reducing conditions [oxygen fugacity, log $fO_2$, ranging from 3-7 units below the iron-wüstite buffer, (Zolotov et al., 2013)]. Based on these characteristics, Mercury's magnetic carriers are likely dominated by the Fe-metal (kamacite) and FeNi alloys (martensite), [see also (Narrett et al., 2025b)], possibly similar to the magnetic mineralogy and properties of the aubrite meteorites [which also are inferred to form in similar extreme reducing conditions, (Rochette et al., 2009)].

Figure 13 shows the range of shock-magnetized volume properties capable of generating a 5 nT crustal field at 20-km altitude. From this, we see that the ~$10^1$ µT amplified antipodal field for Cases 1-5 could be explained by a ~20-km thick layer with an Fe-metal abundance equal to the average total Fe content in the Mercury crust. A ~20 nT field would require a 4✕ higher product of thickness and $\chi_{SRM}$. If the dynamo field was ~10✕ stronger (Case 6) at the time of the Caloris impact (i.e., ~2 µT), the amplified field would have been ~34 µT and could explain these crustal signals with Fe-metal in a magnetized layer just ~7-km thick. As mentioned previously, the Caloris antipode region is found within the H-11 Discovery Quadrangle (22°S-65°S and 270°E-360°E), which is an area of interest defined with widespread ancient (pre-Tolstojan) basins (Spudis & Guest, 1988), fault structures (Watters et al., 2001; Watters & Nimmo, 2009), and chaos terrains (Murray et al., 1974; Trask & Guest, 1975). Relevant for the Hermean magnetization history, these pre-Caloris impact basins (e.g., Andal-Coleridge and Bramante-Schubert) could have implanted more iron-rich materials from the impactor themselves. As such, we also consider that more iron-rich materials (e.g., of chondritic composition) could allow for shallower magnetized layers to create these 5 nT crustal fields, with $\chi_{SRM} \sim 0.01$ requiring layer thicknesses of 1-km to 5-km for amplified fields of ~10-35 µT.

### 4.2 Impact Plasma-SRM in Northern Hemisphere

We now discuss the possibility of the local Hermean field being recorded via SRM in the northern hemisphere during the Caloris impact event. As mentioned previously, plasma compression of the magnetic field within a planetary body can result in ~3✕ amplification of the local field via shielding currents that separate the plasma and surface environs [analogous to the Chapman-Ferraro currents at a planetary magnetopause (Chapman & Ferraro, 1931)]. For the surface in the hemisphere of the impact, the ~3✕ initial field strength amplification for Cases 1-5 is ~1.5 µT and Case 6 is ~6 µT. Based on these field strengths, if there was sufficient pressure to record an SRM, this would require material with $\chi_{SRM} \sim 0.01$ (~10✕ larger than native Hermean Fe-metal) in a >40-km layer to create the ~10-20 nT at 20-km altitude measured crustal fields. SRM in the impact hemisphere is restricted to areas with low heating (so as to not exceed the Curie temperature of the magnetic carrier, which is 780°C for kamacite), and thus would only be possible in locations several basin radii away from the impact center. However, from our global impact simulations, we find that the initial pressure wave (~20 GPa in impact basin) travels faster than the plasma cloud, therefore not allowing for any regions in the impact hemisphere to experience strong pressures (>2 GPa) coinciding with locations of amplified field. Consequently, this means that the most probable location for impact plasma SRM is in the basin antipodal region, where focusing and reflection of pressure waves allows for surface material to record the amplified fields at later times (Figure 11).

Although not explicitly modeled here, we next discuss the possible link between impact plasma amplification and magnetization in the northern hemisphere (the region of low-altitude MESSENGER measurements) produced at the antipodes of basin-forming impacts in the southern hemisphere. We can evaluate this possibility using the results in this study and from previous studies showing a similar plasma magnetic field amplification at the antipode for impacts creating the Imbrium- and Caloris-sized basins (~1,100-km and ~1550-km diameter, respectively). Much of the strong (~10-20 nT at 20 km altitude) measured crustal fields are located between 30°N - 70°N and 90°E - 180°E [Figure 6.19 in (Johnson et al., 2018)]. As such, using the crater catalogs of refs. (Fassett et al., 2012; Orgel et al., 2020), we identify a set of large basins that are found antipodal to this widespan region of strong surface magnetization: Andal-Coleridge (centered at ~41°S/~51°W and with a ~830-km diameter), Matisse-Repin (centered at ~25°S/~75°W and with a ~950-km diameter), Eitkou-Milton (centered at ~23°S/~171°W and with a ~1180-km diameter), and Sadi-Scopus (centered at ~82°S/~44°W center and with ~930-km diameter) [(Fassett et al., 2012; Orgel et al., 2020; Mark A. Wieczorek et al., 2012)]. These large basins are all dated to be pre-Caloris (likely all pre-Tolstojan), and as such, if they produced northern hemisphere magnetization, the resulting crustal fields would have to have survived sequential impact and heating events. Much of the strongly magnetized northern hemisphere region is co-located with 3.9-3.7 Ga old effusive volcanic smooth plains, thought to be associated with or produced by the Caloris-impact event (Wang et al., 2021). Given reasonable estimations of magnetized layer thickness and $\chi_{SRM}$ (Figure 13), it is likely that even if these ancient crustal fields would have survived until the present day, they would constitute a weaker overall magnetic signal compared with TRM generated anomalies. The trade-off in generated crustal field strength for dynamo-recording TRM and impact plasma SRM is the relative higher recording efficiency ($\chi_{TRM} \geq 10\chi_{SRM}$) versus the amplified antipodal field, respectively. As such, this places constraints on the magnetic recording properties of Hermean materials for impact plasma SRM to dominate over TRM. The large impact basin Rembrandt (centered at ~33°S, ~87°E center and with ~700-km diameter) is thought have formed around the time of Caloris (Watters et al., 2009; Mark A. Wieczorek et al., 2012) and is antipodal to a region of weaker ~5-10 nT crustal fields [Figure 6.19 in (Johnson et al., 2018)] that is not co-located with the Caloris-aged effusive volcanism. It is possible that these weak crustal magnetic field signals antipodal to the Rembrandt basin could be (partially) produced by SRM recorded during the impact plasma amplification of the Hermean magnetic field.

**5 Conclusions**

Impact plasma amplification of the Hermean dipole field and IMF can provide a new mechanism for magnetizing the surface. We have demonstrated that the Caloris impact event could magnetize its antipodal region as an SRM recording the impact plasma-amplified local field. We do not find it likely that any significant crustal fields can be recorded in the northern hemisphere during the Caloris-impact plasma amplification process. It is possible that similarly sized basin formation events in the southern hemisphere could have magnetized their antipodes in the northern hemisphere. Yet, we do not find it likely that this process can explain all of the spatially widespread, northern hemisphere surface magnetization measured by the MESSENGER spacecraft (~10-20 nT at 20-km altitude). As the impact plasma process cannot explain all of the strong northern hemisphere crustal magnetic records, we cannot exclude the possibility that a stronger ancient Hermean dynamo sourced the widespread effusive volcanic plains TRM.

Future low altitude measurements by spacecraft like BepiColombo (Heyner et al., 2021; Rothery et al., 2020) of large basin antipodes (e.g., Caloris, Rembrandt, Andal-Coleridge, Matisse-Repin, Eitkou-Milton, and Sadi-Scopus) could provide information on the source field and process for magnetizing the surface, which in turn could help place a relative age (compared to the impact event). For example, high surface resolution measurements of the crustal fields antipodal to Caloris can constrain the age of formation of the chaos terrains (Rodriguez et al., 2020; Schultz & Gault, 1975). Furthermore, future Hermean surface sample return could enable both paleomagnetic and petrologic experiments to determine the history of Mercury's dynamo, test the impact field amplification process, and search for evidence of SRM (Tikoo et al., 2015). Lastly, our study of impact plasma amplification on Mercury provides evidence that this process can be important for explaining surface magnetization on airless terrestrial bodies and should be considered as a source of crustal magnetization.

**Acknowledgments**

ISN thanks Nuno F. Loureiro for his keen insight on the MHD simulations. ISN, BPW, and RO thank the NASA Solar System Workings (grants 80NSSC22K0105) and the NASA FINESST (grant 80NSSC23K1363) programs. KM is fully supported by the Australian Research Council (FT210100063) and Curtin University. We gratefully acknowledge the developers of iSALE-2D (https://isale-code.github.io). Resources supporting this work were provided by the NASA High-End Computing (HEC) Program through the NASA Advanced Supercomputing (NAS) Division at Ames Research Center. The authors acknowledge the MIT SuperCloud and Lincoln Laboratory Supercomputing Center for providing (HPC, database, consultation) resources that have contributed to the research results reported within this paper/report.

**Open Research**

All data needed to evaluate the conclusions in the paper are present in the paper. The BATS-R-US model is open source as part of the Space Weather Modeling Framework (http://github.com/SWMFsoftware). The version of BATS-R-US used for this project can be found on Zenodo: 10.5281/zenodo.14545144. Access to the iSALE code can be requested via https://isale-code.github.io.

**Conflict of Interest**

The authors declare no competing interests.

manuscript submitted to *JGR: Planets*

manuscript submitted to *JGR: Planets*Schield, M. A. (1969). Pressure balance between solar wind and magnetosphere. *Journal of Geophysical Research, 74*(5), 1275-1286. https://doi.org/10.1029/JA074i005p01275

Schultz, P. H., & Gault, D. E. (1975). Seismic effects from major basin formations on the moon and mercury. In *The Moon, 12*(2), 159-177. https://doi.org/10.1007/bf00577875

Spudis, P. D., & Guest, J. E. (1988). Stratigraphy and geologic history of Mercury. In F. Vilas, C. R. Chapman, & M. S. Matthews (Eds.), *Mercury* (pp. 118–164). University of Arizona Press.

Stevenson, D. J. (2003). Planetary magnetic fields. *Earth and Planetary Science Letters, 208*(1-2), 1-11. https://doi.org/10.1016/s0012-821x(02)01126-3

Strauss, B. E., Feinberg, J. M., & Johnson, C. L. (2016). Magnetic mineralogy of the Mercurian lithosphere. *Journal of Geophysical Research: Planets, 121*(11), 2225-2238. https://doi.org/https://doi.org/10.1002/2016JE005054

Strom, R. G., Chapman, C. R., Merline, W. J., Solomon, S. C., & Head, J. W., 3rd. (2008). Mercury cratering record viewed from MESSENGER's first flyby. *Science, 321*(5885), 79-81. https://doi.org/10.1126/science.1159317

Tikoo, S. M., Gattacceca, J., Swanson-Hysell, N. L., Weiss, B. P., Suavet, C., & Cournède, C. (2015). Preservation and detectability of shock-induced magnetization. *Journal of Geophysical Research: Planets, 120*(9), 1461-1475. https://doi.org/10.1002/2015je004840

Tóth, G., van der Holst, B., Sokolov, I. V., De Zeeuw, D. L., Gombosi, T. I., Fang, F., Manchester, W. B., Meng, X., Najib, D., Powell, K. G., Stout, Q. F., Glocer, A., Ma, Y.-J., & Opher, M. (2012). Adaptive numerical algorithms in space weather modeling. *Journal of Computational Physics, 231*(3), 870-903. https://doi.org/10.1016/j.jcp.2011.02.006

Trask, N. J., & Dzurisin, D. (1984). Geologic map of the Discovery (H-11) quadrangle of Mercury. *USGS Misc. Investig. Ser. Map I–1658*. https://doi.org/10.3133/i1658

Trask, N. J., & Guest, J. E. (1975). Preliminary geologic terrain map of Mercury. *Journal of Geophysical Research, 80*(17), 2461-2477. https://doi.org/10.1029/JB080i017p02461

van der Ven, H., Niemann-Tuitman, B. E., & Veldman, A. E. P. (1997). An explicit multi-time-stepping algorithm for aerodynamic flows. *Journal of Computational and Applied Mathematics, 82*(1-2), 423-431. https://doi.org/10.1016/s0377-0427(97)00054-x

Vidotto, A. A. (2021). The evolution of the solar wind. *Living Rev Sol Phys, 18*(1), 3. https://doi.org/10.1007/s41116-021-00029-w

Vidotto, A. A., Gregory, S. G., Jardine, M., Donati, J. F., Petit, P., Morin, J., Folsom, C. P., Bouvier, J., Cameron, A. C., Hussain, G., Marsden, S., Waite, I. A., Fares, R., Jeffers, S., & do Nascimento, J. D. (2014). Stellar magnetism: empirical trends with age and rotation. *Monthly

**Figures and Tables**

| Case | $B_{IMF}$ (nT) | Notes |
|---|---|---|
| Case 1 | (0, 0, 500) | Idealized impact from surface tangential to IMF. This geometry is analogous to the equatorial impact from the lunar dipole case. Solar wind parameters in Table 2. Dipole center is (0,0,0) $R_M$ and of equatorial strength 200 nT (dipole moment ~$3\times10^{19}$ Am$^2$). Antipodal surface normal unit vector given by $\hat{\boldsymbol{n}}_i$. |
| Case 2 | (500, 0, 0) | Idealized impact from surface parallel to IMF. This geometry is analogous to the polar impact from the lunar dipole case. Solar wind parameters in Table 2. Dipole center is (0,0,0) $R_M$ with equatorial strength 200 nT (dipole moment ~$3\times10^{19}$ Am$^2$). Antipodal surface normal unit vector given by $\hat{\boldsymbol{n}}_i$. |
| Case 3 | (250, 0, 433) | Impact at ~30° N (Caloris location) with tangential IMF at impact antipode surface. Solar wind parameters in Table 2. Dipole center is (0,0,0.2) $R_M$ with equatorial strength 200 nT (dipole moment ~$3\times10^{19}$ Am$^2$). Antipodal surface normal unit vector given by $\hat{\boldsymbol{n}}_i$. |
| Case 4 | (433, 0, 250) | Impact at ~30° N (Caloris location) with IMF directed radially inward at impact antipode. Solar wind parameters in Table 2. Dipole center is (0,0,0.2) $R_M$ with equator strength 200 nT (dipole moment ~$3\times10^{19}$ Am$^2$). Antipodal surface normal unit vector given by $\hat{\boldsymbol{n}}_i$. |
| Case 4 Reverse | (-433, 0, -250) | Impact at ~30° N (Caloris location) with IMF directed radially outward at impact antipode. Solar wind parameters in Table 2. Dipole center is (0,0,0.2) $R_M$ and of equator strength 200 nT (dipole moment ~$3\times10^{19}$ Am$^2$). Antipodal surface normal unit vector given by $\hat{\boldsymbol{n}}_i$. |
| Case 5 | (-360, 313, -150) | Impact at ~30° N (Caloris location) with IMF set by Parker spiral angle for ancient Mercury. Solar wind parameters in Table 2. Dipole center is (0,0,0.2) $R_M$ and of equator strength 200 nT (dipole moment ~$3\times10^{19}$ Am$^2$). Antipodal surface normal unit vector given by $\hat{\boldsymbol{n}}_i$. |
| 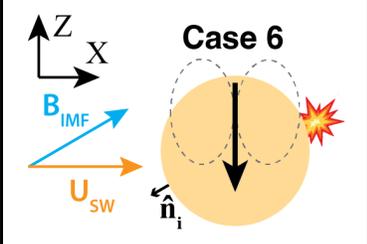 Case 6 | (433, 0, 250) | Like Case 4, but stronger dipole. Impact at ~30° N (Caloris location) with IMF directed radially inward at impact antipode. Solar wind parameters in Table 2. Dipole center is (0,0,0.2) $R_M$ with equatorial strength 2000 nT (dipole moment ~$3\times10^{19}$ Am$^2$). Antipodal surface normal unit vector given by $\hat{\boldsymbol{n}}_i$. |

**Table 1**: Impact plasma simulation cases parameter space. Columns list (left) simulation case, (middle) IMF vector components, $\boldsymbol{B_{IMF}}$, in nT, and (right) notes on the design of the case, impact geometry, and dipole field. Details of the findings for each case can be found in Section 3.

| Plasma Parameters | Solar Wind | Impact Plasma |
|---|---|---|
| Bulk Speed ($u$) | 600 km s$^{-1}$ | 0 |
| Mass density ($\rho$) | 1464 amu cm$^{-3}$ | $4.8 \times 10^{21}$ amu cm$^{-3}$ (surface max.) |
| Magnetic Field ($B$) | 500 nT | Local Magnetospheric Surface Field |
| Temperature ($T$) | 300,000 K | 4000 K |
| Thermal Pressure ($p$) | 6 nPa | $1.1 \times 10^7$ Pa |

**Table 2**: Plasma parameters for the driving boundary conditions. Columns list (left) parameter name, (middle) solar wind value, and (right) impact plasma basin surface boundary condition. The solar wind geometry and other simulation parameters can be found in Table 1. The derivation of the solar wind and impact plasma characteristics can be found in Section 2.

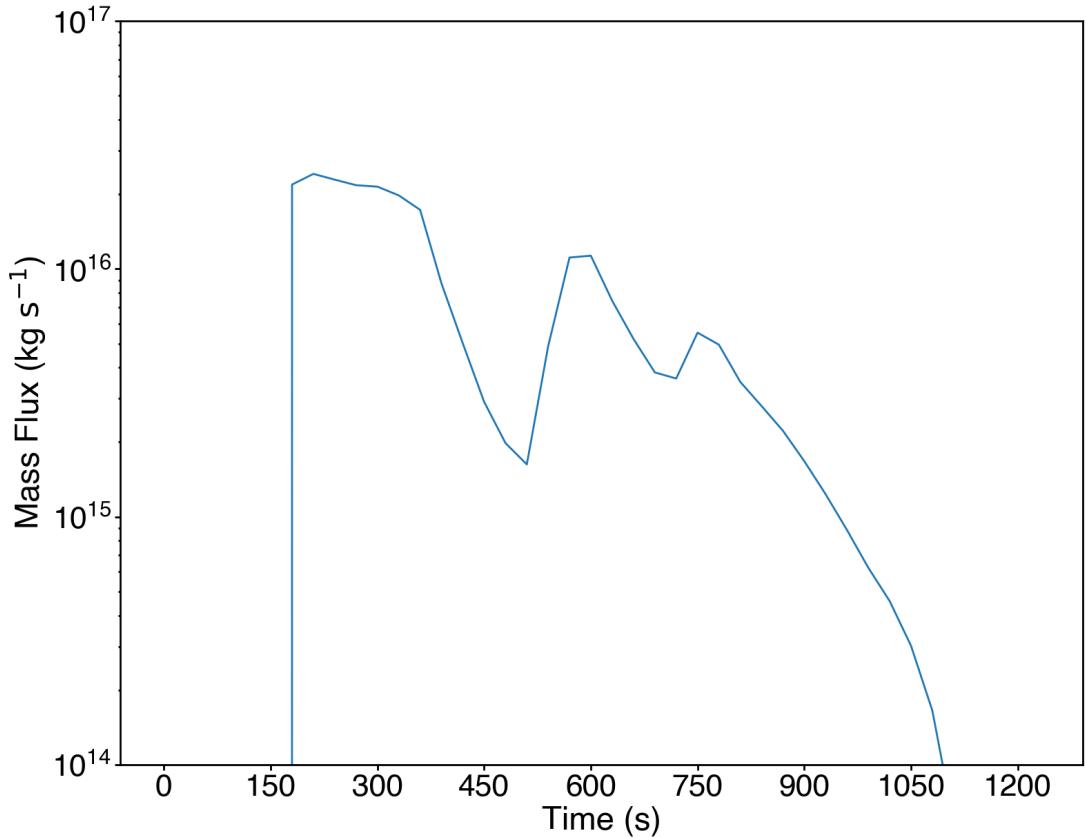

**Figure 1: Impact vapor density flux generated from the Caloris impact event.** Shown is the impact plasma density flux calculated from the iSALE-2D impact simulation of a Caloris-sized impactor. The mass flux shown here was calculated for the vapor (taken to be ionized plasma) with vertical velocity at ~500-km above the surface central impact point, taking the product of the vapor density, velocity, and cross-sectional surface area of a cylindrical volume defined by the crater diameter and ~500-km height. This height was chosen because it captures the density of the impact plasma that has sufficient vertical velocity to expand out of the forming crater and interact with the Mercury magnetic field environment. The calculated density flux agrees well with the analytical estimation of the total vapor mass produced for this impact event (see Section 2).

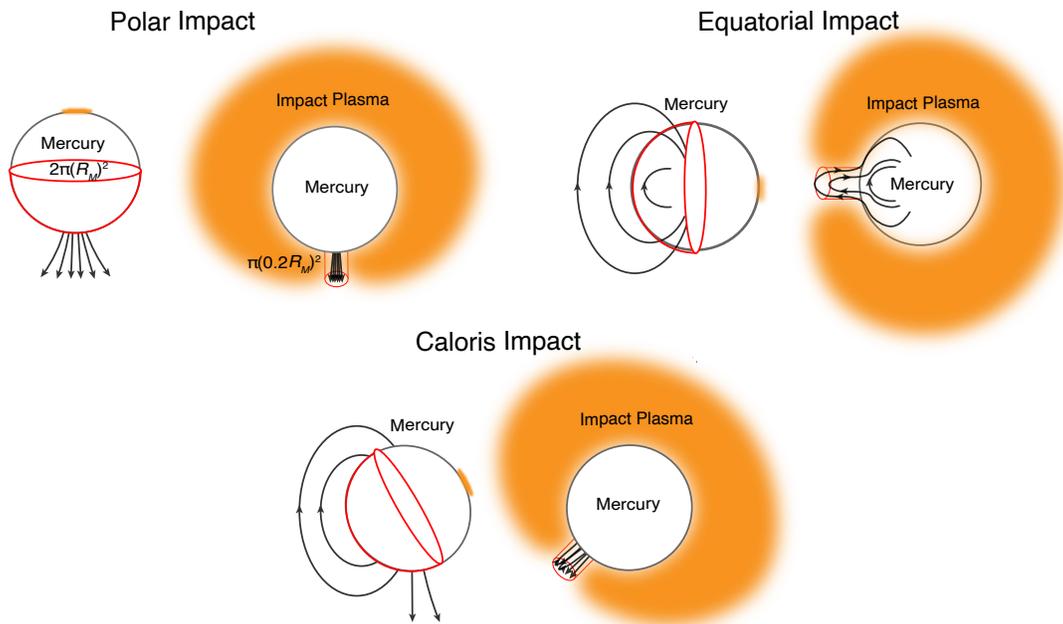

**Figure 2: Cartoon depiction of the dipole magnetic field amplification process for varying impact location.** Each set of cartoons consists of two images depicting the (left) initial state impact plasma (orange "cloud") expansion and (right) stage of maximum magnetic field amplification for impacts that occur at the magnetic pole [Polar impact (Narrett et al., 2025a)], the magnetic equator [Equatorial impact (Narrett et al., 2025a)], and the ~30°N Caloris location (Caloris impact, Cases 3, 4, 4 – Reverse, 5, and 6). The red hemispherical outline in each set of images (left) shows the initial surface area through which the internally-generated dipole magnetic field (black arrows) threads, while the cylindrical volume (right) shows the compressed magnetic field geometry. The polar impact results in the maximum magnetic field seen in these scenarios, because the initially parallel field lines at the antipode are compressed into a smaller volume and therefore higher flux (i.e., the change of area through which magnetic field threads determines field strength change). The Equatorial impact (top right) results in the smallest amplified field due to the compression of initially anti-parallel field within the compressed volume. The Caloris impact (bottom) partially consists of equatorial field lines, meaning that the final compressed geometry contains some fraction of initially anti-parallel field lines, resulting in a diminished total amplification relative to the Polar impact. From previous studies (Narrett et al., 2025a), impacts at ~30° from the magnetic equator resulted in ~11× amplification compared to the maximum ~21× amplification from the magnetic pole.

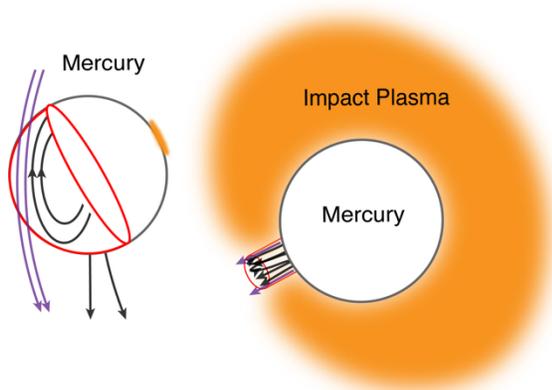 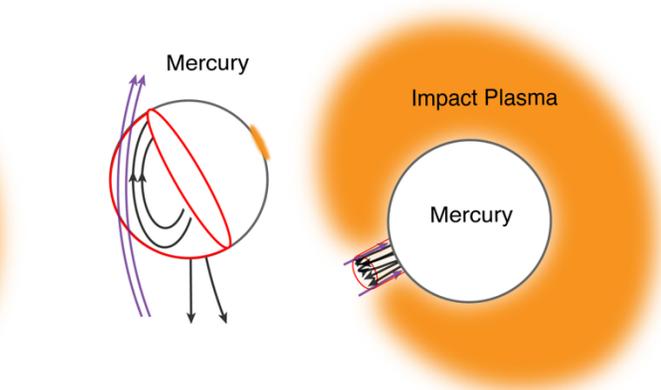

**Figure 3: Cartoon depiction of the magnetic field amplification process for changing IMF geometry.** Each set of cartoons consists of two images depicting the (left) initial state impact plasma (orange "cloud") expansion and (right) stage of magnetic field amplification for impacts with the IMF and dipole field ("Parallel") parallel and ("Anti-parallel") anti-parallel in the Caloris antipode region. The red hemispherical outline in each set of images (left) shows the initial surface area through which the internally-generated dipole (black arrows) and IMF (purple arrows) thread while the cylindrical volume (right) shows the compressed magnetic field geometry. The (left) parallel dipole and IMF geometry results in greater amplification when compared to (right) anti-parallel dipole and IMF geometry, due to the cancelling effect of the oppositely aligned field. The anti-parallel geometry amplification (Case 3 and Case 4) can be approximated with the ~21× maximum amplification factor (Narrett et al., 2025a) multiplied by the average antipodal surface field and then subtracting the ~11× amplification of the oppositely aligned dipole field (Narrett et al., 2025a) for impacts at Caloris location. This parallel geometry amplification (Case 4 – Reverse) can be approximated with the ~21× maximum amplification factor multiplied by the average antipodal surface field.

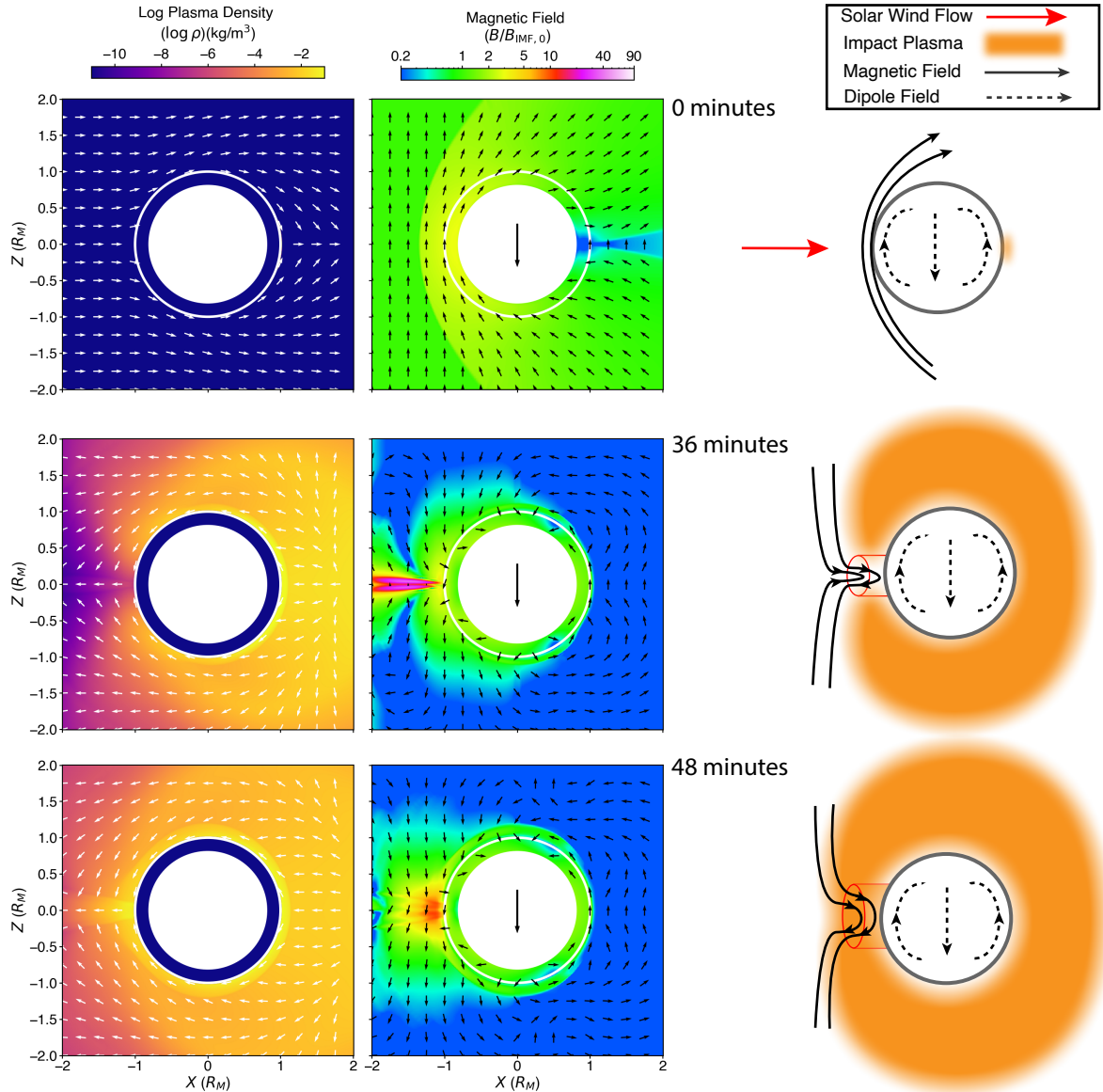

**Figure 4: Amplification of the IMF and Hermean dynamo field by a Caloris-sized impact when the IMF is perpendicular to the impact antipode normal (Case 1).** 2D slices of the 3D-MHD simulation of impact plasma expanding from the surface (~200 nT surface equatorial field, body-centered dipole) magnetic equator within the ancient Hermean environment. The top row shows the initial condition steady-state Hermean magnetic field environment, while the middle and bottom rows show the impact plasma expansion at 36 and 48 minutes after impact, respectively; these are the times of approximate maximum antipodal surface magnetic field of ~3 µT (blue curves in Figure 10) and magnetic field relaxation, respectively. The left column shows the evolution of the (log-scale) plasma mass density, $\rho$, with velocity flow direction (white arrows). The middle column shows the evolution of the total magnetic field magnitude normalized to the 0.5 µT IMF, $B/B_{IMF,0}$, with magnetic field direction (black arrows). The right column shows a cartoon depiction of the impact plasma (orange "cloud") expanding and compressing the superposition of the IMF and Hermean dipole field (black arrowed streamlines) into the antipodal region. The middle panel illustrates the magnetic geometry, by which the impact plasma compresses anti-parallel field lines together into a small region, increasing the

magnitude of the magnetic field. The bottom panel shows a widened cylindrical circular face, representing the relaxation of the amplified magnetic field due to the further expansion of the impact plasma and the resistive Hermean surface, which dissipates magnetic field energy. The white circular outline depicts the Hermean surface while the shaded white region depicts the (~0.8$R_M$) core surface. The dashed black arrows show the internal dipole field geometry generated from the Hermean dynamo. The solar wind is flowing in the *+X* direction with characteristics described in Tables 1 and 2. The impact is launched from (*X*=1, *Z*=0 $R_M$).

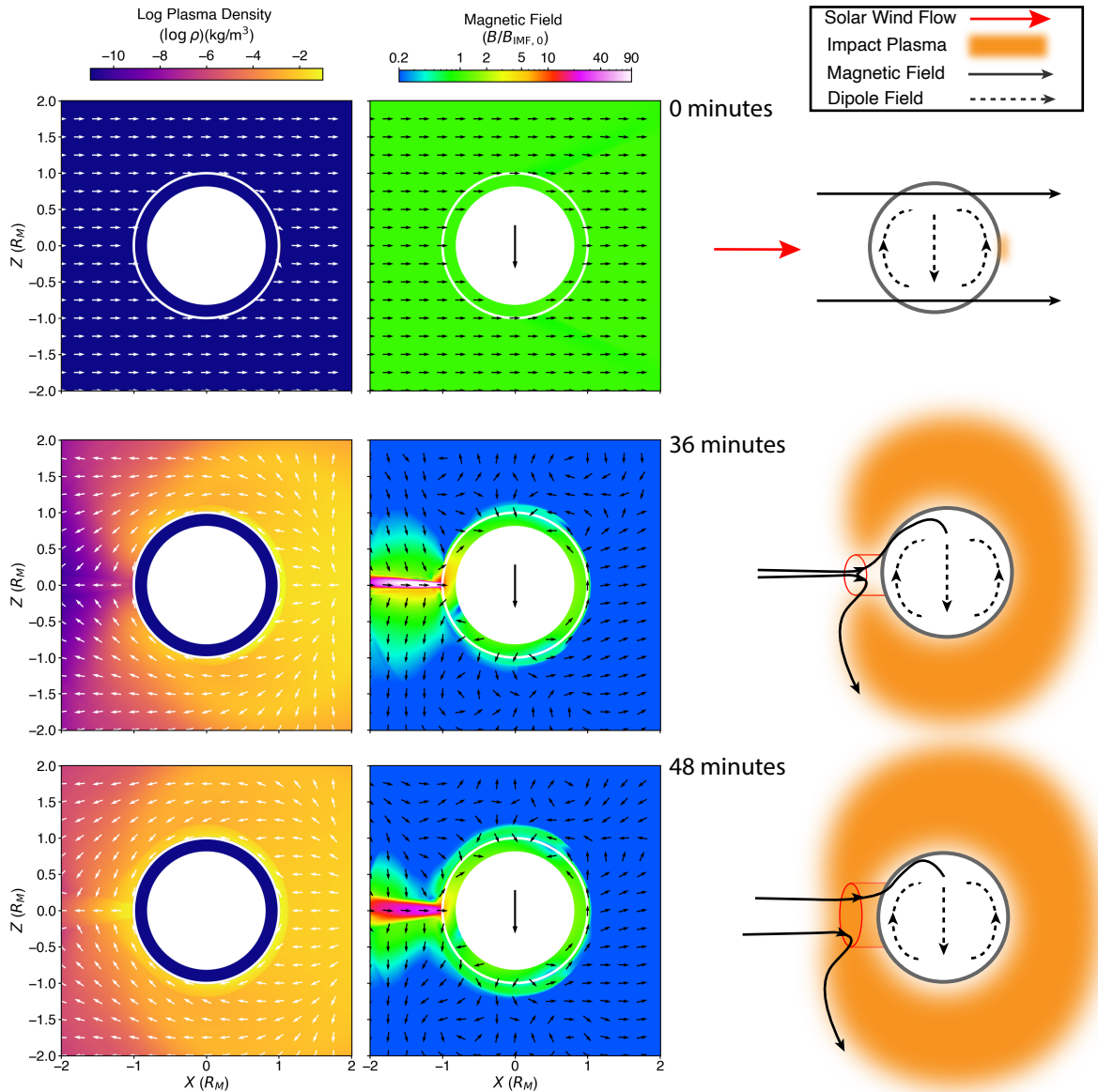

**Figure 5: Amplification of the IMF and Hermean dynamo field by a Caloris-sized impact when the IMF is parallel to the impact antipode normal (Case 2).** 2D slices of the 3D-MHD simulation of impact plasma expanding from the surface (~200 nT, body-centered dipole) magnetic equator within the ancient Hermean environment. The top row shows the initial condition steady-state Hermean magnetic field environment, while the middle and bottom rows show the impact plasma expansion at 36 and 48 minutes after impact, respectively; these are the times of approximate maximum antipodal surface magnetic field of ~11 µT (orange curves in Figure 10) and magnetic field relaxation, respectively. The left column shows the evolution of the (log-scale) plasma mass density, $\rho$, with velocity flow direction (white arrows). The middle column shows the evolution of the total magnetic field magnitude normalized to the 0.5 µT IMF, $B/B_{\text{IMF},0}$, with magnetic field direction (black arrows). The right column shows a cartoon depiction of the impact plasma (orange "cloud") expanding and compressing the superposition of the IMF and Hermean dipole field (black arrowed streamlines) into the antipodal region. The middle row cartoon illustrates the magnetic geometry, by which the impact plasma compresses

parallel field lines together into a small region, maximizing the magnetic field amplification. The bottom row cartoon shows a widened cylindrical circular face, representing the relaxation of the amplified magnetic field due to the further expansion of the impact plasma and the resistive Hermean surface, which dissipates magnetic field energy. The white circular outline depicts the Hermean surface while the shaded white region depicts the (~$0.8R_M$) core surface. The dashed black arrows show the internal dipole field geometry generated from the Hermean dynamo. The solar wind is flowing in the *+X* direction with characteristics described in Tables 1 and 2. The impact is launched from (*X=1, Z=0 $R_M$*).

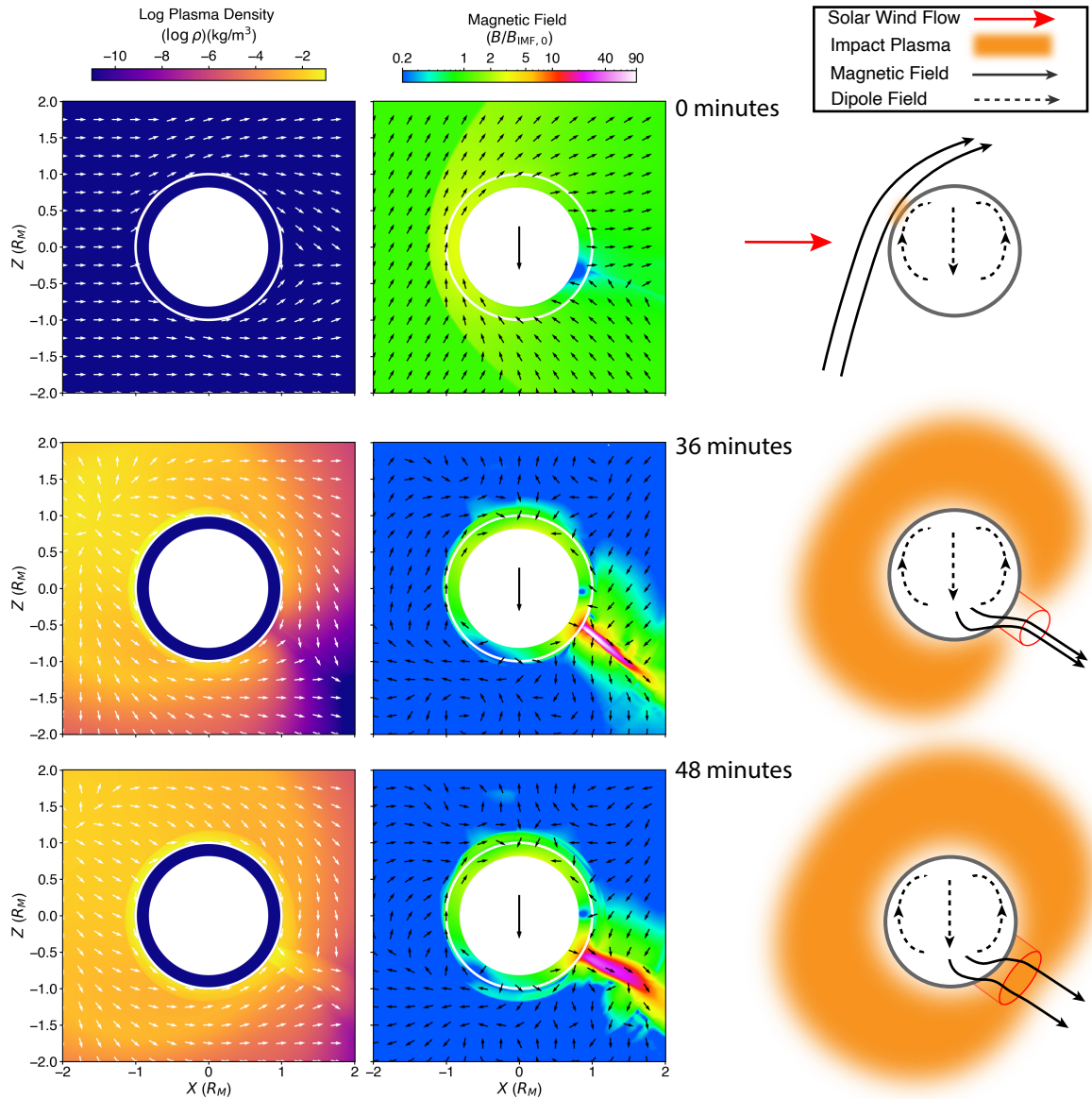

**Figure 6: Amplification of the IMF and Hermean dynamo field by a Caloris-sized impact at ~30°N when the IMF is perpendicular to the impact antipode normal (Case 3).** 2D slices of the 3D-MHD simulation of impact plasma expanding from ~30°N (~200 nT, center-shifted dipole) within the ancient Hermean environment. The top row shows the initial condition steady-state Hermean magnetic field environment, while the middle and bottom rows show the impact plasma expansion at 36 and 48 minutes after impact, respectively; these are the times of approximate maximum antipodal surface magnetic field of ~13.7 µT (green curves in Figure 10) and magnetic field relaxation, respectively. The left column shows the evolution of the (log-scale) plasma mass density, $\rho$, with velocity flow direction (white arrows). The middle column shows the evolution of the total magnetic field magnitude normalized to the 0.5 µT IMF, $B/B_{IMF,0}$, with magnetic field direction (black arrows). The right column shows a cartoon depiction of the impact plasma (orange "cloud") expanding and compressing the superposition of the IMF and Hermean dipole field (black arrowed streamlines) into the antipodal region. The middle row cartoon illustrates the magnetic geometry, by which the impact plasma compresses

parallel field lines (superposition of parallel IMF and dipole field) together into a small region, maximizing the magnetic field amplification. The bottom row cartoon shows a widened cylindrical circular face, representing the relaxation of the amplified magnetic field due to the further expansion of the impact plasma and the resistive Hermean surface, which dissipates magnetic field energy. The white circular outline depicts the Hermean surface while the shaded white region depicts the (~$0.8R_M$) core surface. The dashed black arrows shows the internal dipole field geometry generated from the Hermean dynamo shifted northward from the planet's center by ~$0.2R_M$ (Anderson et al., 2011). The solar wind is flowing in the $+X$ direction with characteristics described in Tables 1 and 2. The impact is launched from ($X = -0.866$, $Z = 0.5$ $R_M$).

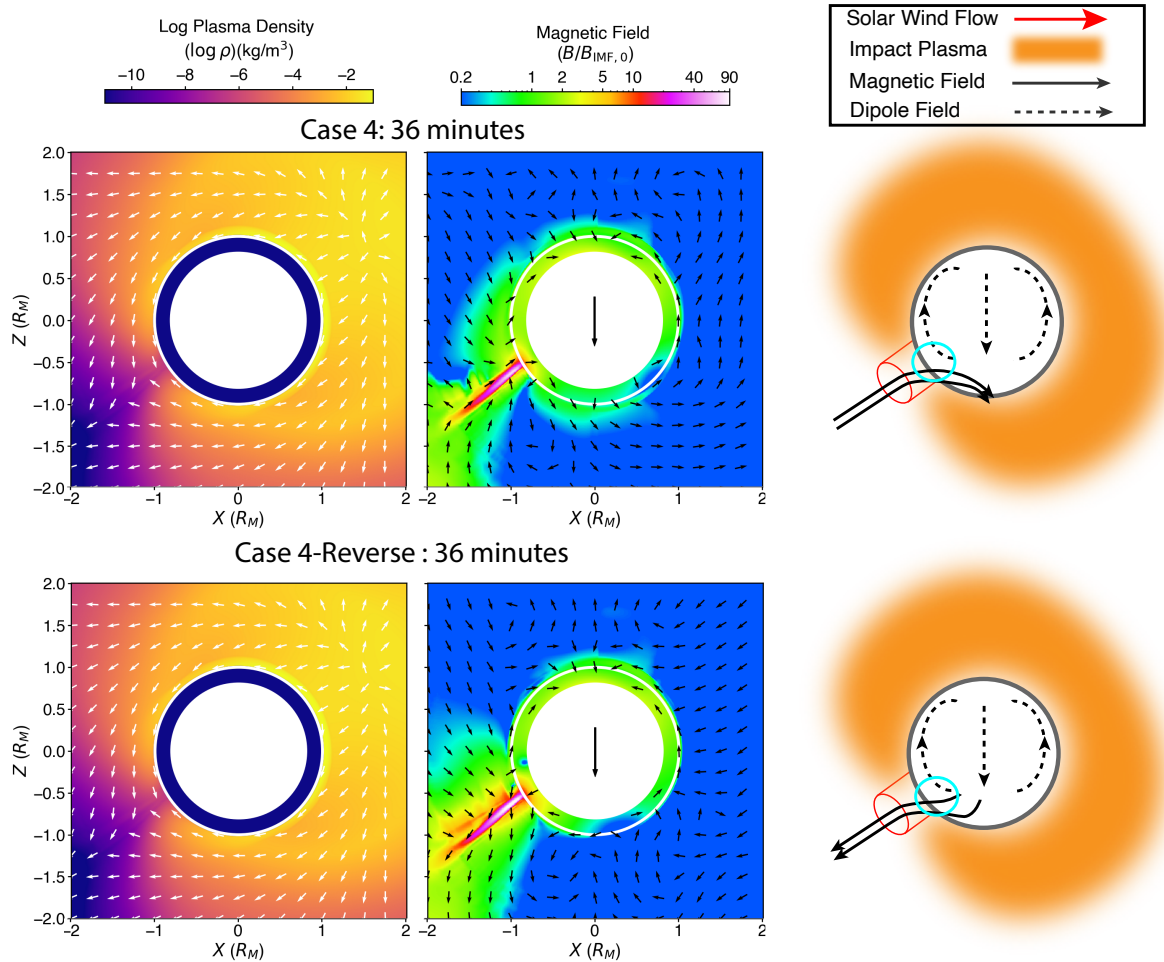

**Figure 7: Amplification of the IMF and Hermean dynamo field by a Caloris-sized impact at ~30°N when the IMF is parallel and anti-parallel to the impact antipode normal (Cases 4 and 4-Reverse).** 2D slices of the 3D-MHD simulation of impact plasma expanding from ~30°N (~200 nT, center-shifted dipole) within the ancient Hermean environment. The rows show the impact plasma expansion at 36 minutes for (top) Case 4 and (bottom) 4-Reverse; these are the times of approximate maximum antipodal surface magnetic field of ~10 µT and ~12.7 µT (red and purple curves in Figure 10), respectively. The left column shows the evolution of the (log-scale) plasma mass density, $\rho$, with velocity flow direction (white arrows). The middle column shows the evolution of the total magnetic field magnitude normalized to the 0.5 µT IMF, $B/B_{\mathrm{IMF},0}$, with magnetic field direction (black arrows). The right column shows a cartoon depiction of the impact plasma (orange "cloud") expanding and compressing the superposition of the IMF and Hermean dipole field (black arrowed streamlines) into the antipodal region. In both cases, the antipodal convergence region sees a strongly amplified magnetic field due to the compression of parallel magnetic field lines. However, Case 4-Reverse results in a higher total magnetic field due to the parallel IMF and dipole field geometry, whereas these fields are anti-parallel in Case 4 (seen in the blue circular outline, see Figure 3). The white circular outline depicts the Hermean surface while the shaded white region depicts the (~$0.8R_M$) core surface. The dashed black arrows shows the internal dipole field geometry generated from the Hermean dynamo shifted northward from the planet's center by ~$0.2R_M$ (Anderson et al., 2011). The solar

wind is flowing in the +*X* direction with characteristics described in Tables 1 and 2. The impact is launched from (*X* = 0.866, *Z* = 0.5 $R_M$).

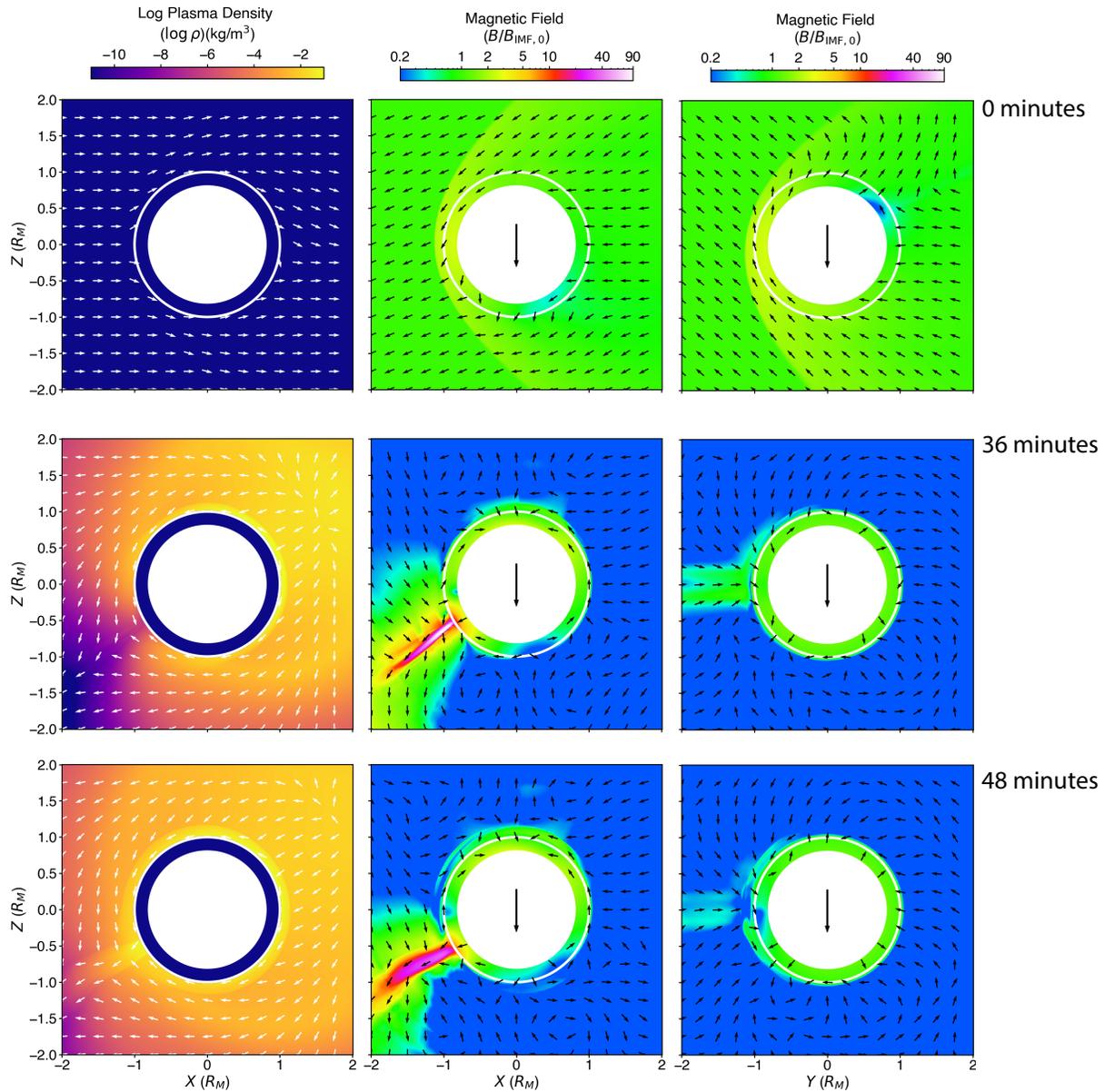

**Figure 8: Amplification of the IMF and Hermean dynamo field by a Caloris-sized impact at ~30°N when the IMF has an planet-orbital direction component (Case 5).** 2D slices of the 3D-MHD simulation of impact plasma expanding from ~30°N (~200 nT, center-shifted dipole) within the ancient Hermean environment. The top row shows the initial condition steady-state Hermean magnetic field environment, while the middle and bottom rows show the impact plasma expansion at 36 and 48 minutes after impact, respectively; these are the times of approximate maximum antipodal surface magnetic field of ~14 μT (brown curves in Figure 10) and magnetic field relaxation, respectively. The left column shows the evolution of the (log-scale) plasma mass density, $\rho$, with velocity flow direction (white arrows). The middle and right columns show the evolution of the total magnetic field magnitude normalized to the 0.5 μT IMF, $B/B_{IMF,0}$, with magnetic field direction (black arrows) in the *X-Z* and *Y-Z* planes, respectively. The white circular outline depicts the Hermean surface while the shaded white region depicts the (~$0.8R_M$) core surface. The dashed black arrows shows the internal dipole field geometry generated from the Hermean dynamo shifted northward from the planet's center by ~$0.2R_M$

(Anderson et al., 2011). The solar wind is flowing in the $+X$ direction with characteristics described in Tables 1 and 2. The impact is launched from ($X = 0.866$, $Z = 0.5$ $R_M$).

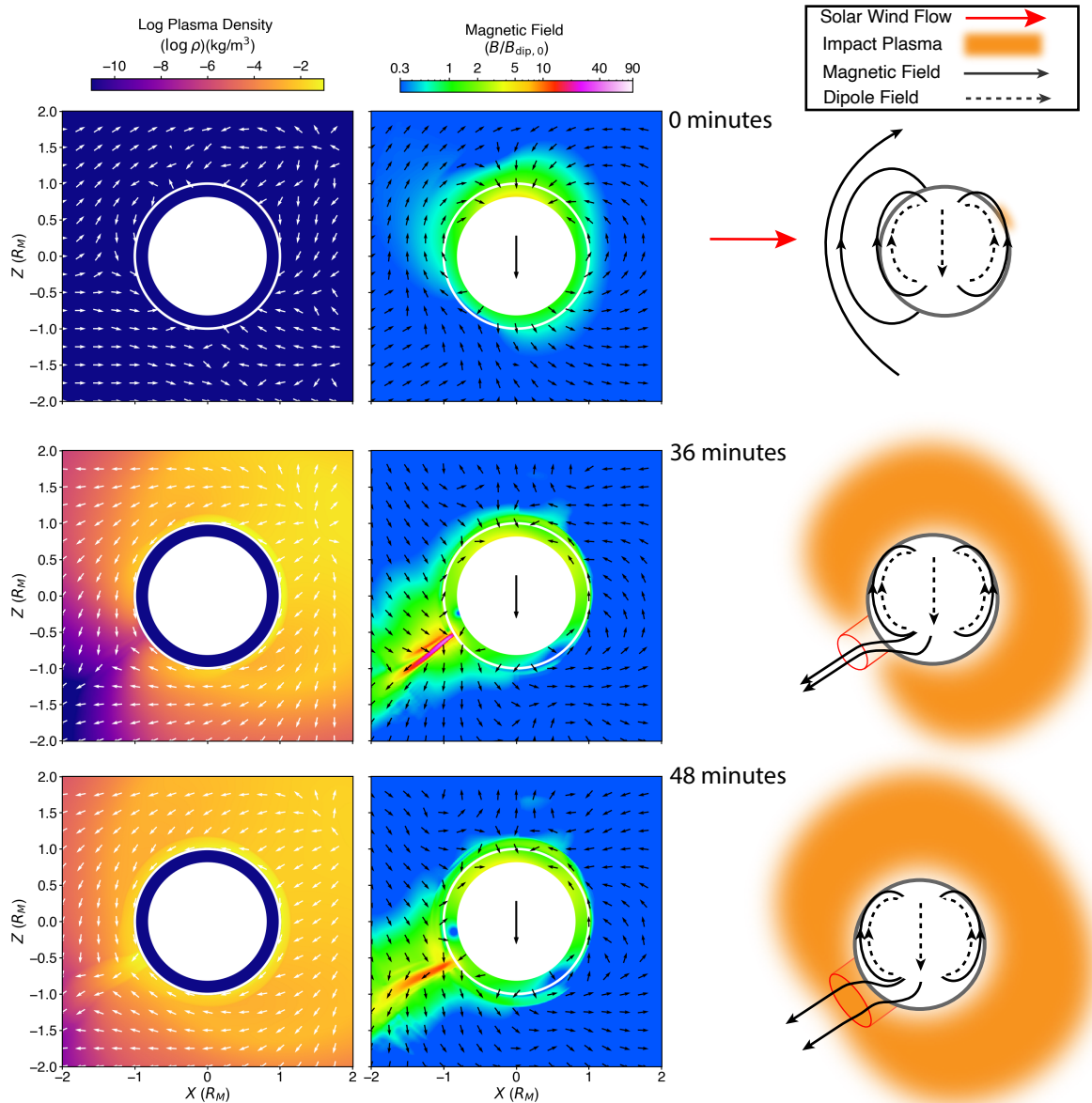

**Figure 9: Amplification of the IMF and 10× stronger Hermean dynamo field by a Caloris-sized impact at ~30°N (Case 6).** 2D slices of the 3D-MHD simulation of impact plasma expanding from ~30°N (~2 μT, center-shifted dipole) within the ancient Hermean environment. The top row shows the initial condition steady-state Hermean magnetic field environment, while the middle and bottom rows show the impact plasma expansion at 36 and 48 minutes after impact, respectively; these are the times of maximum antipodal surface magnetic field of ~33 μT (pink curves in Figure 10) and magnetic field relaxation, respectively. The left column shows the evolution of the (log-scale) plasma mass density, $\rho$, with velocity flow direction (white arrows). The middle column shows the evolution of the total magnetic field magnitude normalized to the 2 μT equatorial dipole field, $B/B_{\text{dip},0}$, with magnetic field direction (black arrows). The right column shows a cartoon depiction of the impact plasma (orange "cloud") expanding and compressing the superposition of the IMF and Hermean dipole field (black arrowed streamlines) into the antipodal region. The middle row cartoon illustrates the magnetic geometry, by which the impact plasma compresses parallel field lines (superposition of parallel IMF and dipole field)

together into a small region, maximizing the magnetic field amplification. The bottom row cartoon shows a widened cylindrical circular face, representing the relaxation of the amplified magnetic field due to the further expansion of the impact plasma and the resistive Hermean surface, which dissipates magnetic field energy. The white circular outline depicts the Hermean surface while the shaded white region depicts the (~$0.8R_M$) core surface. The dashed black arrows shows the internal dipole field geometry generated from the Hermean dynamo shifted northward from the planet's center by ~$0.2R_M$ (Anderson et al., 2011). The solar wind is flowing in the *+X* direction with characteristics described in Tables 1 and 2. The impact is launched from (*X*=0.866, *Z*=0.5 $R_M$).

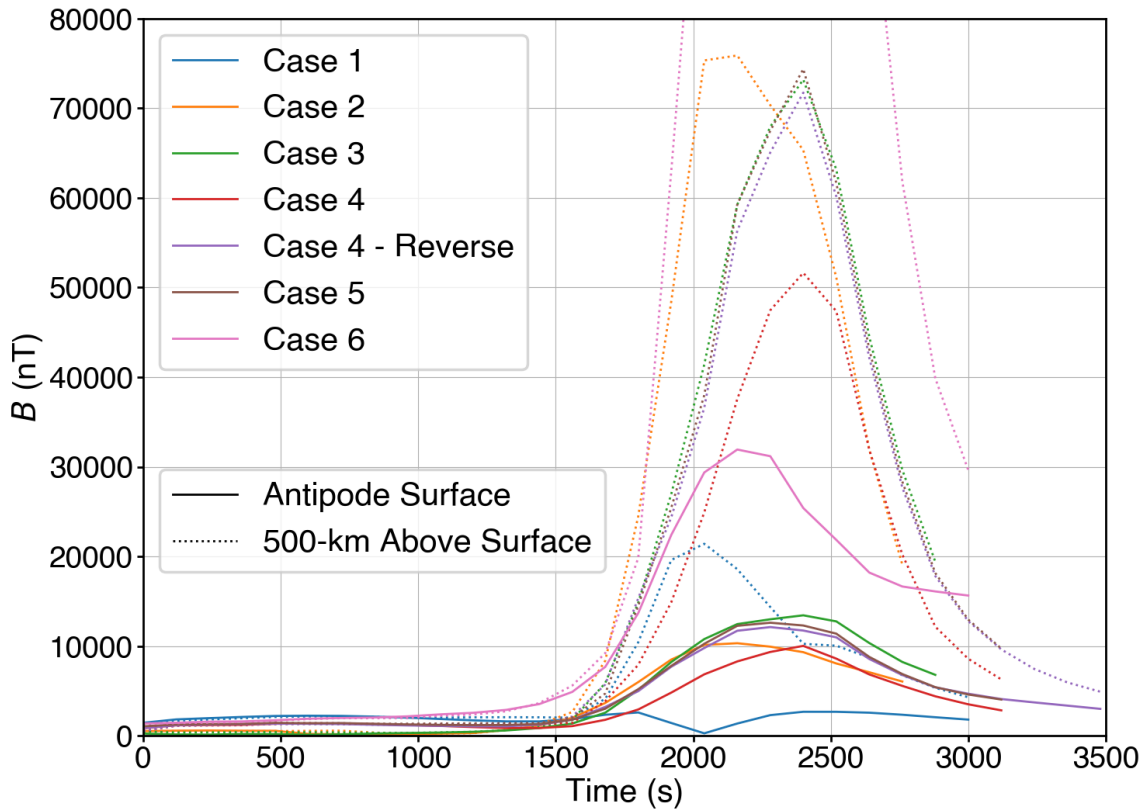

**Figure 10: Antipodal surface and near-surface magnetic field evolution during the expansion of impact plasma**. Line plots depicting the time evolution of the antipodal (solid) surface and (dashed) ~500-km altitude (above surface) magnetic field magnitude for all cases. Cases 1 and 2 represent idealized IMF, impact location, and dipole field geometries, shown to result in (Case 1) lesser and (Case 2) maximized antipodal magnetic field due to the compression of anti-parallel and parallel magnetic field, respectively. Cases 3, 4, 4-Reverse, and 5 account for the Caloris impact location and the modern shifted dipole-center, resulting in similar amplified magnetic fields of ~10-14 µT that occur between 35 and 40 minutes. Case 6 shows a heightened final amplified magnetic field of ~33 µT due to the initial ~10× stronger dipole. From our global impact simulations (Figure 11), we find strong pressure waves (>0.4 GPa) that occur in the antipodal crust during this time-window, enabling acquisition of the impact-plasma amplified field in the surface material via SRM.

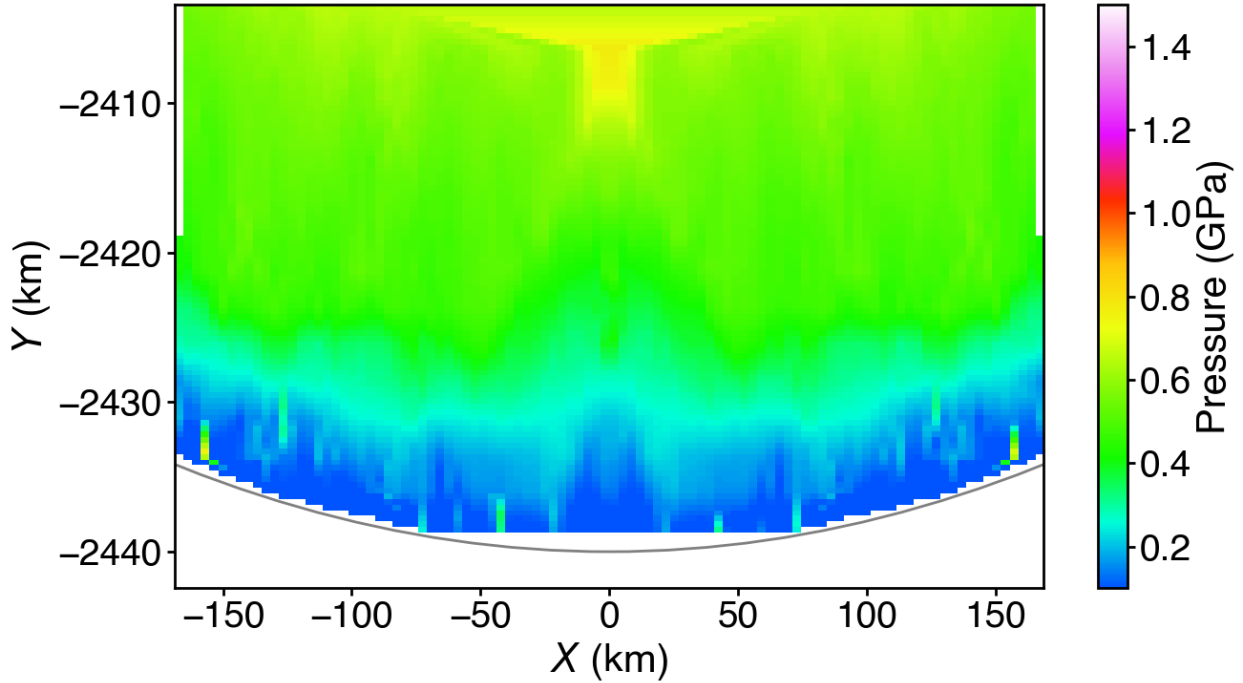

**Figure 11: Impact simulation showing maximum strength of body pressure wave convergence in the antipode.** Pressure in the antipodal region starting at time of impact (and lasting until ~6,000 seconds) from the iSALE-2D impact simulation of a Caloris-basin sized event used to derive the impact plasma parameters for the 3D-MHD simulation. Spatial distribution of maximum pressure experienced within the ~150-km radius surface area and ~40-km depth antipodal region during times of amplified magnetic field (>10 µT). This calculation shows that the antipodal crust experiences pressures >0.2 GPa and averages to ~0.4 GPa, consistent with previous predictions (Hughes et al., 1977; Schultz & Gault, 1975; Watts et al., 1991), enabling the acquisition of SRM at the time of maximum antipodal field amplification.

manuscript submitted to *JGR: Planets*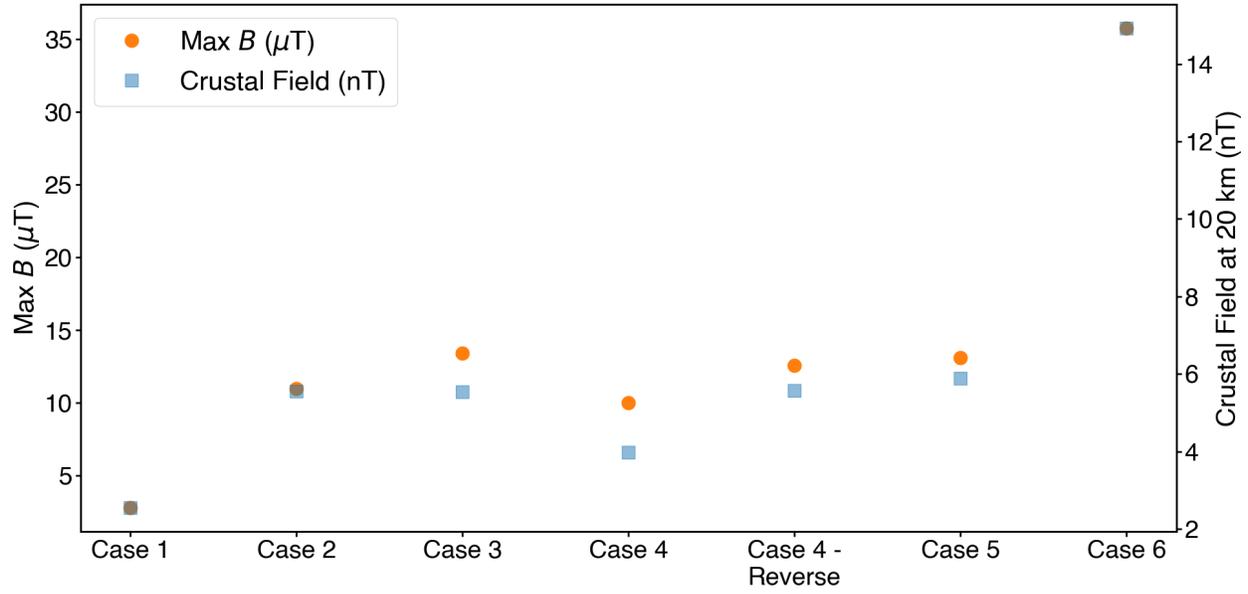

**Figure 12: Maximum antipodal magnetic field and generated crustal field from impact-induced SRM.** Summary plot of the (left vertical axis, orange circles) maximum impact plasma amplified magnetic field and (right vertical axis, blue squares) generated crustal field at 20-km altitude, typical for low-altitude passes made by MESSENGER (Johnson et al., 2018). The generated crustal fields are calculated using an SRM recording efficiency, $\chi_{SRM}= 0.01$, and using an analytical expression for a uniformly magnetized cylinder of radius 20-km and thickness 10-km (Caciagli et al., 2018). The generated crustal field is proportional to $\chi_{SRM}$, so these values can be scaled for other materials as shown in Figure 13.

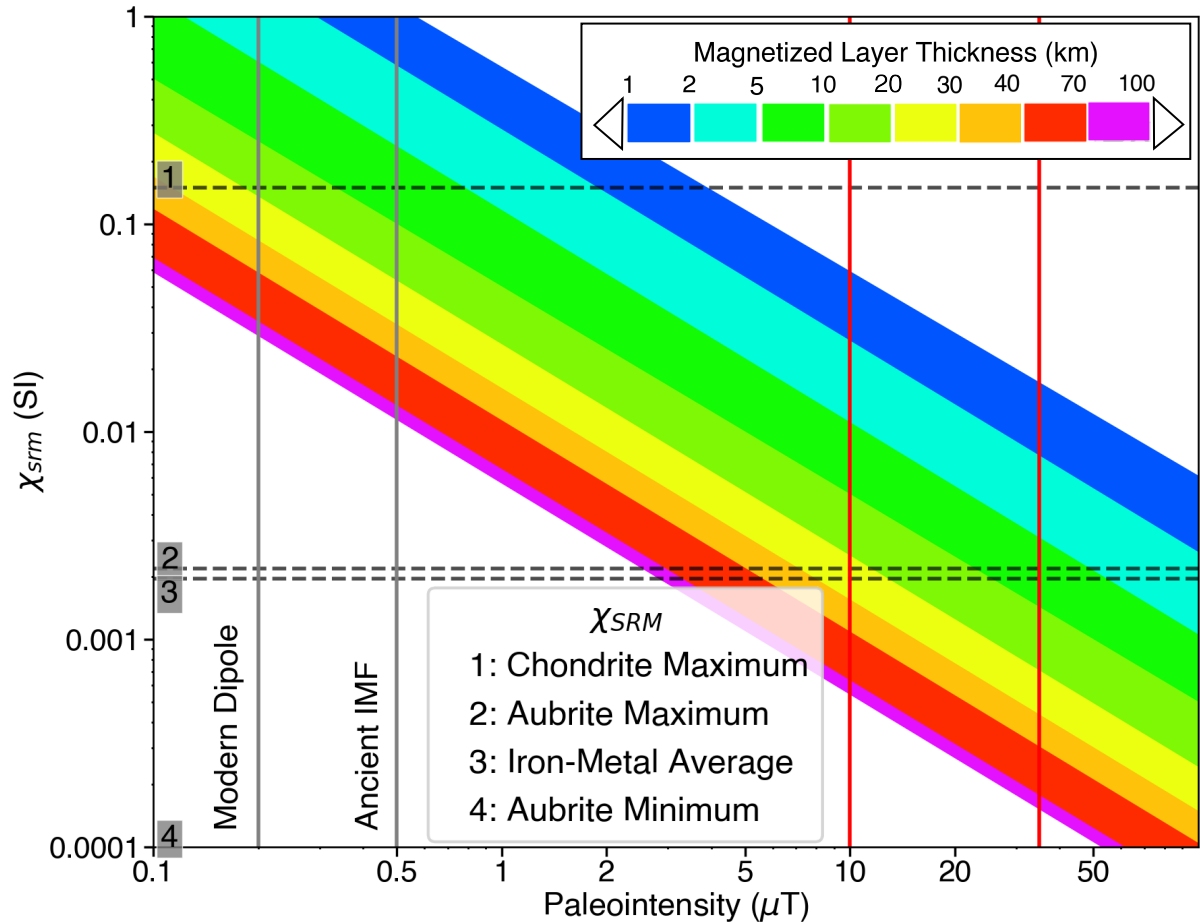

**Figure 13: Shock remanent magnetization efficiency, paleointensity, and layer thickness of magnetized materials that can produce 5 nT crustal anomalies.** Thicknesses (color bar) required to magnetize crustal material and generate a 5 nT field observed at 20-km altitude for varying ambient, ancient magnetic fields (paleointensities) and SRM efficiencies ($\chi_{SRM}$). The magnetized material is derived to represent a cylindrical disk volume with a 20-km radius and ranging thicknesses (Caciagli et al., 2018). The $\chi_{SRM}$ values are calculated from the mean published $\chi_{TRM}$ values in (Kletetschka et al., 2006; Rochette et al., 2009; M. A. Wieczorek et al., 2012) for possible native Hermean material (aubrite-like or iron-metal) and non-native, impact delivered materials (chondritic composition) representing an upper limit with $\chi_{SRM} = 0.1\chi_{TRM}$, in line with the SRM experiments of (Tikoo et al., 2015). The grey vertical lines represent the initial magnetic field strengths, with the leftmost being the 0.2 µT dipole field and rightmost being the 0.5 µT ancient IMF. The red vertical lines represent the range of maximum amplified surface fields (~10-33 µT) that can be recorded by the surface material from impact-induced pressure wave SRM. From this paleointensity range, it is possible that ~5 nT crustal fields could be generated with $\chi_{SRM} \sim 0.01$ for layers of thickness between 2 and 10 km.